\documentclass[12pt,letterpaper]{article}
\usepackage[a4paper, total={7in, 10in}]{geometry}

\usepackage{graphicx}
\usepackage{helvet}
\usepackage{authblk}
\usepackage{hyperref}
\usepackage{amsmath} 
\usepackage{amssymb} 
\usepackage{eurosym}
\usepackage{orcidlink} 
\usepackage{multirow}
\usepackage{booktabs}
\usepackage[super,comma,sort&compress]  
   {natbib}
\usepackage[right]{lineno} 

\makeatletter
\renewcommand{\maketitle}{\bgroup\setlength{\parindent}{0pt}
\begin{flushleft}
  \textbf{\@title}
  
  \@author
\end{flushleft}\egroup}
\makeatother

\title{EU-ETS under attack? The impact of carbon price suppression on the decarbonization of the power sector
}
\date{}

\author[1,2,3,5,*,\orcidlink{0009-0000-5794-406X}]{Javier Gonzalez-Ruiz}
\author[1,2,3,**]{Carlos Rodriguez-Pardo}
\author[1,2,3]{Alice Di Bella}
\author[4]{Paolo Mastropietro}
\author[4]{Jose Pablo Chavez-Avila.}
\author[1,2,3]{Massimo Tavoni}

\affil[1]{Politecnico di Milano, Piazza Leonardo da Vinci 32, Milan, 20133, Italy}
\affil[2]{CMCC Foundation- Euro-Mediterranean Center on Climate Change, Via Marco Biagi 5, Lecce, 73100, Italy}
\affil[3]{RFF-CMCC European Institute on Economics and the Environment, Via Bergognone 34, Milan, 20144, Italy}
\affil[4]{Instituto de Investigacion Tecnologica, Universidad Pontificia Comillas, Rey Francisco 4, Madrid, Spain}
\affil[5]{Lead contact}

\affil[*]{Correspondence: javier.gonzalez@cmcc.it}
\affil[**]{Correspondence: carlos.rodriguezpardo.jimenez@gmail.com}

\begin{document}

\maketitle

\section*{SUMMARY}

European countries are debating policies to mitigate the increased energy costs caused by renewed geopolitical tensions, while pursuing decarbonization and electrification. A notable example is Italy's 2026 \textit{Decreto Bollette} package, which proposes to remove the carbon price equivalent from the bids of certain gas-driven power plants to wholesale electricity markets, among other provisions. We use this as a case study to assess the long-term implications of suppressing the carbon price signal in the electricity market for investment, emissions, and consumer costs. We employ a stylized Italian power system using MARLEY, a multi-agent reinforcement learning framework focused on long-term electricity market assessments. In this framework, we test this policy across configurations with varying levels of support for green investment, resource adequacy, and flexibility. Results show that partial suppression of the carbon price signal yields short-term cost reductions but only a minor long-term effect on total system costs, as the deferred emissions are ultimately repaid by consumers. CO$_2$ Emissions rise across most configurations since suppressing the price signal erodes incentives for renewable and storage investment. Only the most ambitious configurations for supporting green investment avoid this outcome, but they do so by marginalizing the wholesale price signal itself, thereby requiring a commitment to a hybrid market paradigm that is in contradiction with the rationale of the proposed price intervention. 

\section*{KEYWORDS}


Electricity Market Design, Hybrid Markets, Multi-Agent Reinforcement Learning, Decarbonization Pathways, EU ETS, Carbon Pricing

\clearpage

\section*{INTRODUCTION}
\label{Section - INTRODUCTION}

In recent years, large-scale geopolitical events have put recurring pressure on electricity prices, as in 2022, after the intensification of the Russo-Ukrainian war, and more recently with the 2026 Iran war and the consequent closure of the Strait of Hormuz \cite{cambridge_middle_east_and_north_africa_forum_2026_2026}. High electricity prices are putting the energy transition in Europe's power sectors under strain. Governments are forced to decide how to reduce consumer bills without undermining long-term climate policy goals, while also reducing Europe's structural dependence on fossil fuel imports \cite{batlle_power_2022,batlle_power_2022-1}. In response to these crises, the EU and its Member States launched the REPowerEU plan and adopted a range of fiscal and structural measures to shield households and businesses from surging energy costs \cite{european_commission_repowereu_2022}. 

At the national level, some governments intervened directly in wholesale market pricing. In 2022, Spain and Portugal implemented the \textit{Iberian exception} \cite{jefatura_del_estado_espana_real_2022}, a temporary compensation to fossil-fired generators financed through a surcharge on consumers. Generators were required to internalize this compensation in their market bids, thus reducing the marginal price. While the \textit{Iberian exception} achieved considerable reductions in spot electricity prices \cite{haro_ruiz_effects_2024,fabra_winners_2025}, it also created significant economic inefficiencies, particularly with regard to cross-border trade, and increased the perceived regulatory risk \cite{linares_assessment_2023,fabra_unpacking_2025}. Facing analogous pressures, the Italian government adopted a comparable measure in 2026: the \textit{Decreto Bollette} \cite{repubblica_italiana_misure_nodate}. Among its provisions, the Decree proposes to compensate fossil-fired generators for the carbon price signal embedded in the European Union Emissions Trading System (EU ETS), lowering the marginal cost of price-setting gas plants and thereby reducing the clearing price and the inframarginal rents of dispatched generators \cite{visconti_parisio_decreto_2026}. To recover the EU ETS quantities owed, the decree establishes a specific charge in end-user tariffs \footnote{We acknowledge that this mechanism may be modified, or not activated at all, following the European Commission analysis on state-aid clearance. Nonetheless, provided that any revisions to the policy package do not substantially alter the carbon price suppression itself, the effects would remain aligned with those described in this work.}

The partial neutralization of the carbon price signal from electricity price formation raises fundamental questions about the coherence of these policies with the decarbonization objective, widely recognized as the cornerstone of a broader economic transformation \cite{intergovernmental_panel_on_climate_change_working_2022,iea_world_2024}, with Renewable Energy Sources (RES) and energy storage systems (ESS) becoming the central building blocks of future power systems \cite{iea_world_2024,irena_world_2024}. This transition is based on two pillars, and the suppression of the carbon price interacts directly with both.

The first pillar concerns the role of carbon pricing as a long-term decarbonization instrument. A large body of empirical work has documented that carbon costs borne by power producers are passed through to wholesale electricity prices \cite{sijm_co2_2006,fabra_pass-through_2014,bai_drivers_2023}. This pass-through raises consumer prices, but also generates the investment signals that support cleaner technologies by widening the cost differential between fossil and low-carbon assets \cite{lilliestam_effect_2021}. As such, carbon pricing also partially mitigates the risk of cannibalization faced by renewable investors \cite{hirth_market_2013} by increasing the perceived remuneration of these assets \cite{hirth_market_2013,pena_cannibalization_2022}. On the other hand, by making electricity more expensive, it risks delaying the adoption of electric technologies which are needed to decarbonize the end-use sectors and increase energy efficiency.

The second pillar is the growing role of long-term regulatory mechanisms in supporting decarbonization. Capacity Remuneration Mechanisms (CRM), support schemes for RES, storage or demand response, and other contracting schemes have become widely implemented programs to de-risk investment, improve resource adequacy, and shield consumers from price volatility \cite{us_department_of_energy_benefits_2006,oren_generation_2005}. In the last two decades, these regulatory instruments have become pivotal for decarbonization, facilitating capital-intensive investment, mitigating missing-money and missing-market problems, and addressing the specific challenges of RES-based systems \cite{cramton_capacity_2013,newbery_missing_2016}. In the European Union, Contract for Difference (CfD) auctions, complementing Power Purchase Agreements (PPAs) and forward markets, have become the dominant instrument for achieving mitigation targets in the electricity sector \cite{zachmann_design_2023,newbery_efficient_2023,schlecht_financial_2024,european_parliament_improving_2024}. Importantly, mechanism design has shifted away from decoupling support incentives from wholesale signals (first generation feed-in-tariffs) toward instruments that couple directly to short-term price formation, so their performance now depends on whether the carbon price signal is included in the dispatch.

\begin{figure}[hbt!]
\centering
\includegraphics[width=1.0\linewidth]{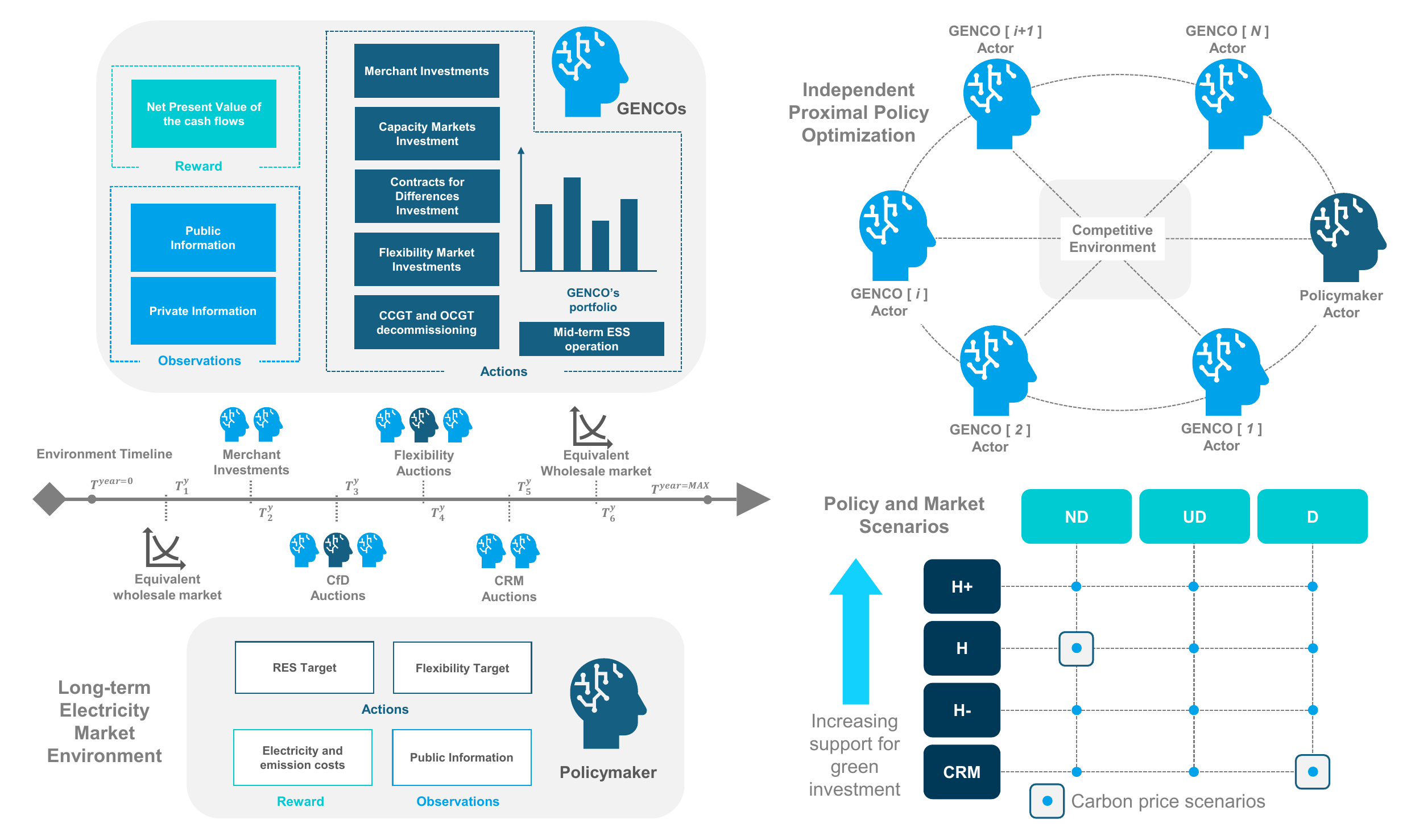} 
\caption{\textbf{Structure of the long-term electricity market in the reinforcement learning model.} The left panel presents the market environment, highlighting the agent's observations, actions, and rewards, as well as the interactions in the equivalent wholesale market and the investment mechanisms. The upper-right panel introduces Independent Learning for Proximal Policy Optimization. Lastly, the lower-right panel shows the scenario matrix combining policy and market conditions, along with the carbon price scenarios applied to specific cases.}
\label{fig: General Overview}
\end{figure}

In this context, carbon price suppression, the pace and resilience of decarbonization, the design and implementation of long-term regulatory mechanisms, and the impacts on final consumers are tightly coupled: the suppression alters investment incentives, which reshape the generation mix and its CO$_{2}$ emissions, which in turn change the policy response and the prices ultimately borne by consumers. Capturing this feedback within a single modeling exercise is precisely the challenge we address, aiming to understand the implications of measures analogous to the \textit{Decreto Bollette} and the \textit{Iberian exception} for decarbonization pathways in power systems. To achieve this objective, we follow the general framework presented in Figure \ref{fig: General Overview}.

To start, we use the Multi-Agent Reinforcement Learning for Electricity markets (MARLEY) framework applied to a stylized representation of the Italian electricity system \cite{gonzalez-ruiz_assessing_2026}. MARLEY is an open-source agent-based model, based on Independent Proximal Policy Optimization (IPPO), dedicated to analyzing the evolution of electricity markets under a wide range of long-term mechanisms to promote investment. Recent advances in deep learning have extended multi-agent reinforcement learning (MARL) to highly complex competitive domains \cite{openai_dota_2019,vinyals_grandmaster_2019,yu_surprising_2022}, including applications to electricity markets spanning short-term bidding strategies \cite{ye_multi-period_2019,ye_deep_2020,du_approximating_2021,graf_computational_2023,harder_fit_2023}, peer-to-peer markets \cite{qiu_mean-field_2023}, and multi-agent frameworks incorporating active regulators whose policy instruments co-evolve endogenously with market actors' strategies \cite{renshaw-whitman_non-stationarity_2024}. MARLEY draws on both agent-based \cite{chappin_simulating_2017,anwar_can_2024,barazza_co-evolution_2020} and partial-equilibrium models \cite{gabriel_complementarity_2014,billimoria_insurance_2022,mays_missing_2021,dimanchev_choosing_2024}, the two dominant approaches for electricity market modeling \cite{bublitz_survey_2019}, while complementing and extending them in four directions: (i) it explicitly incorporates long-term regulatory mechanisms, (ii) it integrates policy layers such as carbon pricing, (iii) it allows agents to manage portfolios of new and existing assets, and (iv) it captures strategic interactions in concentrated wholesale markets.

The explicit integration of a representative policymaker into a MARL model studied by \citet{renshaw-whitman_non-stationarity_2024} is particularly relevant to our work, as a suppressed carbon price creates a feedback loop between policymakers' decisions on the need for long-term regulatory mechanisms and the investment responses of profit-maximizing Generation Companies (GENCOs). Building on this, whereas the original MARLEY models GENCOs alone \cite{gonzalez-ruiz_assessing_2026}, we add a representative policymaker who, under decarbonization pressure, uses auctions for RES and ESS to shape the evolution of the generation mix. GENCOs, the agents responsible for investing in the electricity market, react to system conditions, market prices, and the actions of other agents, including the policymaker. This design allows us to evaluate the effects of the carbon price signal (and its suppression) in a context in which the policymaker endogenously responds to the distortion it creates.
 
Using MARLEY, we are able to represent the suppressed carbon price signal in the power sector and the provisions of the \textit{Decreto Bollette}, which have the most direct implications for long-term market dynamics. In particular, we focus on the suppression for combined-cycle gas turbine (CCGT) plants, an efficient fossil-fuel technology whose marginal cost most directly governs wholesale price formation in many European power systems. Additionally, we study a 2025–2040 horizon, capturing both the short-term suppression of the carbon price signal and its longer-term investment consequences. For the analysis, Italy is the natural case: the \textit{Decreto Bollette} is the most explicit European instance of removing the carbon price signal from the formation of wholesale prices, and the country's gas-dominated marginal generation makes the mechanism's effects particularly easy to identify \cite{terna_terna_2026}. However, the findings of our analysis are not only applicable to the Italian system; they can also be generalized to any system that might introduce similar interventions on bids from marginal generators using fossil fuels. 

To further characterize the effects of this type of measure, we combine policy scenarios, market design configurations, and selected carbon price trajectories, as shown in the lower-right panel of Figure \ref{fig: General Overview}:
\begin{itemize}
    \item The policy scenarios span three cases. Two are core to the analysis: a baseline without the Decree \textbf{[ND]} and the Decree in place \textbf{[D]}, in which the ETS price is neutralized by exempting natural gas power plants, compensated by an increase in the tariff system \footnote{Although the scenario labels refer to whether the \textit{Decreto Bollette} is applied, we remind the reader that, in this work, its application is represented solely through the suppression of the carbon price signal.}. Moreover, we assume that carbon price neutralization remains in place throughout the horizon considered (until 2040). The third scenario, \textbf{[UD]}, introduces the Decree immediately at the beginning of the simulation horizon, followed by an uncertain timing for the restoration of the full carbon price signal, capturing the implications of an uncertain duration of the \textit{Decreto Bollette}. 
    \item The market design configurations vary in the long-term regulatory mechanisms layered on top of the energy market. The \textbf{[CRM]} scenario supplements it with a capacity remuneration mechanism based on a Reliability Option design, while the hybrid configurations \textbf{[H-]}, \textbf{[H]}, and \textbf{[H+]} build on the CRM by adding production-based CfDs for RES and a dedicated flexibility support mechanism for short-term storage. These hybrid scenarios span a wide range of ambition, measured by the speed of penetration and the maximum size of each mechanism. In particular, the CfDs target renewable penetration of up to 60, 80, and 120\% of average yearly demand in the \textbf{[H-]}, \textbf{[H]}, and \textbf{[H+]} configurations, respectively; the flexibility mechanisms aim for the same short-term energy storage penetration levels. This range approximates the hybrid market paradigm increasingly adopted across European jurisdictions, where the sizing of such mechanisms is itself a policy and political choice tied to the incentives shaping the electricity mix. We therefore treat these configurations as progressively more ambitious benchmarks. Systems resembling current European practice are best represented by \textbf{[CRM]} and \textbf{[H-]}, where long-term mechanisms support but do not displace merchant investment as the primary entry channel. Instead, \textbf{[H]} and \textbf{[H+]} represent systems that increasingly rely on these mechanisms, substantially limiting the role of short-term price signals in investment decisions.    
    \item Finally, we measure the effect of the \textit{Decreto Bollette} across different carbon price trajectories, applied to the \textbf{[CRM]} and \textbf{[H]} scenarios, which serve as proxies for systems with varying degrees of reliance on long-term support for green investments.
\end{itemize}

Following this framework, we contribute to the current policy discussions on the effects of carbon price suppression in electricity markets that rely on long-term regulatory mechanisms. Although focused on the Italian electricity system, the analysis speaks to a broader class of cases: electricity systems simultaneously planning long-term regulatory mechanisms to support the low-carbon transition and facing short-term political pressure to shield consumers from fossil-fuel-driven wholesale price spikes. Specifically, we: (i) quantify the expected total system cost reductions and potential increments in CO$_2$ emissions following the application of interventions intended to reduce the marginal price in the market and their interaction with existing long-term regulatory mechanisms; (ii) assess how carbon price suppression affects long-term investment signals for green technologies; and (iii) characterize the effects of policy uncertainty surrounding price interventions analogous to the \textit{Decreto Bollette}. Alongside these empirical contributions, the paper extends open-source MARL simulation frameworks for electricity market design, with the most updated version to be made publicly available upon acceptance.

\clearpage

\section*{RESULTS}

\begin{figure}[hbt!]
\centering
\includegraphics[width=1.0\linewidth]{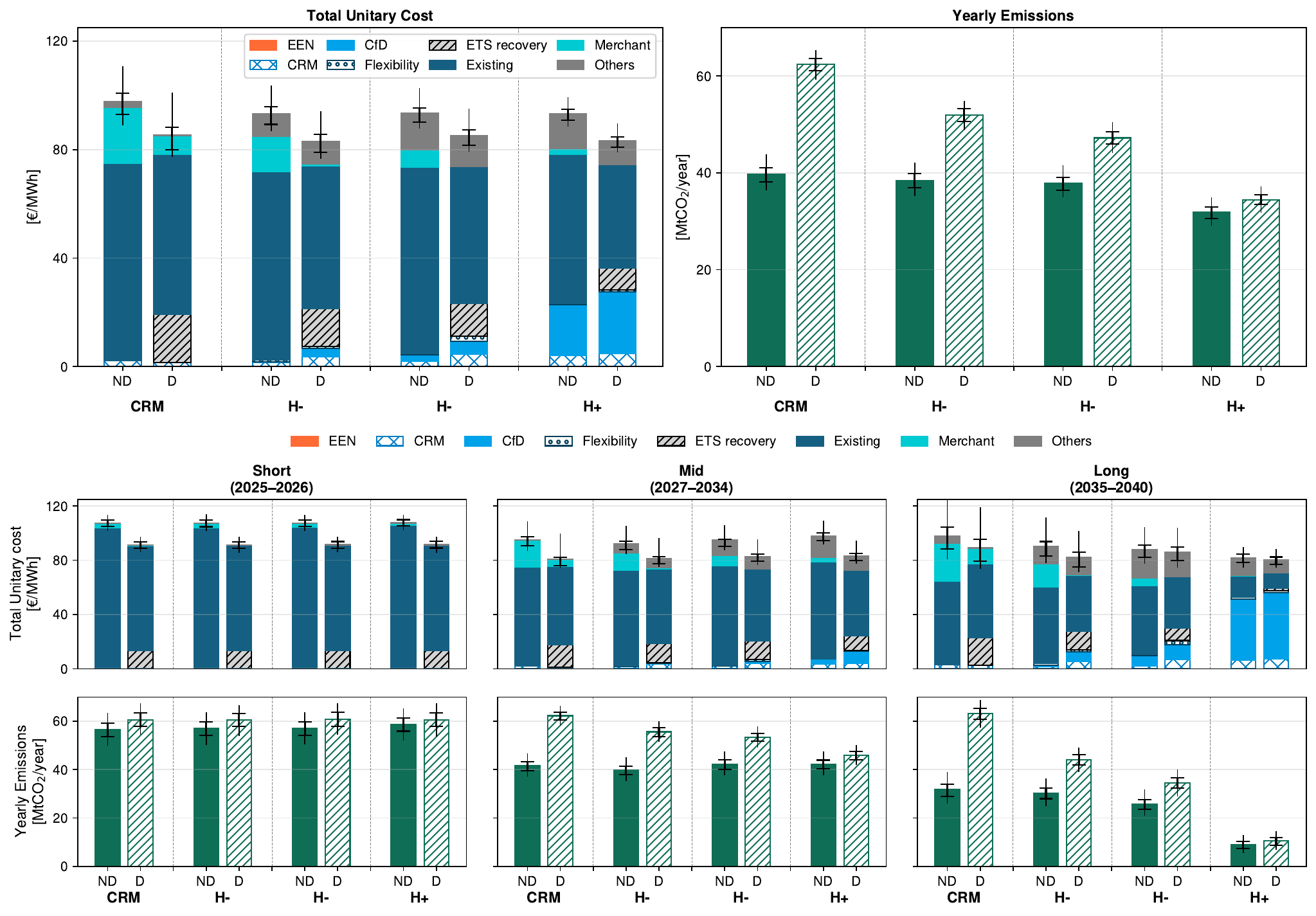}
\caption{\textbf{Total Unitary costs and CO$_2$ emissions in \textbf{[D]} and \textbf{[ND]} scenarios.} The upper panel presents aggregated results for the 2025-2040 horizon, while the lower panel disaggregates them into short-, mid-, and long-term components. In each bar, the horizontal markers display the 25th and 75th percentiles, and the vertical markers display the 5th and 95th percentiles. In the total unitary cost bars, hatching patterns indicate the contribution of each market mechanism to the final cost: the Contracts for Differences, Capacity Market, and Flexibility components reflect the financial settlement of their respective mechanisms; Other captures the wholesale remuneration of capacity and flexibility assets, which remain exposed to the short-term price signal; Existing and Merchant bars refer to the wholesale market remuneration of old and new assets; and ETS recovery represents the ETS costs recovered through the tariff in the \textbf{[D]} scenarios.}
\label{fig: Prices and emissions ND - D}
\end{figure}

Figure \ref{fig: Prices and emissions ND - D} and Table \ref{table: Prices and emissions ND - D} present the aggregate effects on total system costs and emissions for the \textbf{[D]} and \textbf{[ND]} scenarios across market configurations. Results reveal a fundamental trade-off: the partial elimination of the carbon price signal reduces costs in the short-term, but at the expense of increased emissions, the magnitude of which depends on the available long-term regulatory instruments. The lower panels in Figure \ref{fig: Prices and emissions ND - D} highlight the asymmetry between the short- and long-run consequences of suppressing the carbon price. In the short run, wholesale prices decline, reducing generators' revenues. However, total system costs, which incorporate the ETS obligations incurred by CCGT plants and recovered through end-user tariffs, narrow this gap, moderating the benefit perceived by final consumers. As the horizon extends to the mid- and long-run, the dominant effect shifts from price reduction to emissions accumulation. Market configurations without or with low green investment support (\textbf{[H-]} and \textbf{[CRM]} scenarios) show lower emissions reduction over time in the \textbf{[D]} cases, unlike the persistent reductions observed in scenarios with higher support (\textbf{[H]} and \textbf{[H+]}). These observations are corroborated by formal significance tests on the per-simulation outcome distributions presented in the Appendix. The increase in emissions induced by the \textit{Decreto Bollette} is statistically significant across all market configurations and timeframes, with a large effect size in aggregate, placing it clearly outside the simulation uncertainty range. The cost reduction is likewise significant, but corresponds to a markedly smaller effect size that diminishes over the horizon.

\begin{table}[hbt!]
\centering
\resizebox{\textwidth}{!}{%
\begin{tabular}{cclllll}
\hline
\multirow{2}{*}{\textbf{Window}} &
  \multirow{2}{*}{\textbf{Market}} &
  \multicolumn{1}{c}{\textbf{System cost}} &
  \multicolumn{1}{c}{\textbf{Emissions}} &
  \multicolumn{1}{c}{\multirow{2}{*}{\textbf{Market}}} &
  \multicolumn{1}{c}{\textbf{System cost}} &
  \multicolumn{1}{c}{\textbf{Emissions}} \\
 &
   &
  \multicolumn{1}{c}{\textit{{[}\%{]}}} &
  \multicolumn{1}{c}{\textit{{[}\%{]}}} &
  \multicolumn{1}{c}{} &
  \multicolumn{1}{c}{\textit{{[}\%{]}}} &
  \multicolumn{1}{c}{\textit{{[}\%{]}}} \\ \hline
\textbf{Aggregate} & \multirow{4}{*}{CRM} & -11.94 & 56.92 & \multirow{4}{*}{H-} & -10.74 & 35.05 \\
\textbf{Short}     &                      & -14.98 & 7.1   &                     & -15.01 & 5.92  \\
\textbf{Mid}       &                      & -14.37 & 49.88 &                     & -11.27 & 39.65 \\
\textbf{Long}      &                      & -7.58  & 98.83 &                     & -8.31  & 45.25 \\ \hline
\textbf{Aggregate} & \multirow{4}{*}{H}   & -8.8   & 24.6  & \multirow{4}{*}{H+} & -10.6  & 8.16  \\
\textbf{Short}     &                      & -14.94 & 6.76  &                     & -15.1  & 3.23  \\
\textbf{Mid}       &                      & -11.79 & 26.51 &                     & -14.96 & 8.68  \\
\textbf{Long}      &                      & -1.96  & 33.57 &                     & -1.62  & 15.57 \\ \hline
\end{tabular}%
}
\caption{\textbf{Relative percentage difference in total system cost and CO$_2$ emissions between [ND] and [D] scenarios across time horizons.} Values are computed as $(\mathrm{D}-\mathrm{ND})/\mathrm{ND}$: a negative figure indicates that suppressing the carbon price \textbf{[D]} lowers the indicator relative to the \textbf{[ND]} baseline, a positive figure that raises it. Results are reported for the aggregate 2025-2040 horizon and its short-, mid-, and long-term sub-periods, across market configurations.}
\label{table: Prices and emissions ND - D}
\end{table}

\begin{figure}[hbt!]
\centering
\includegraphics[width=1.0\linewidth]{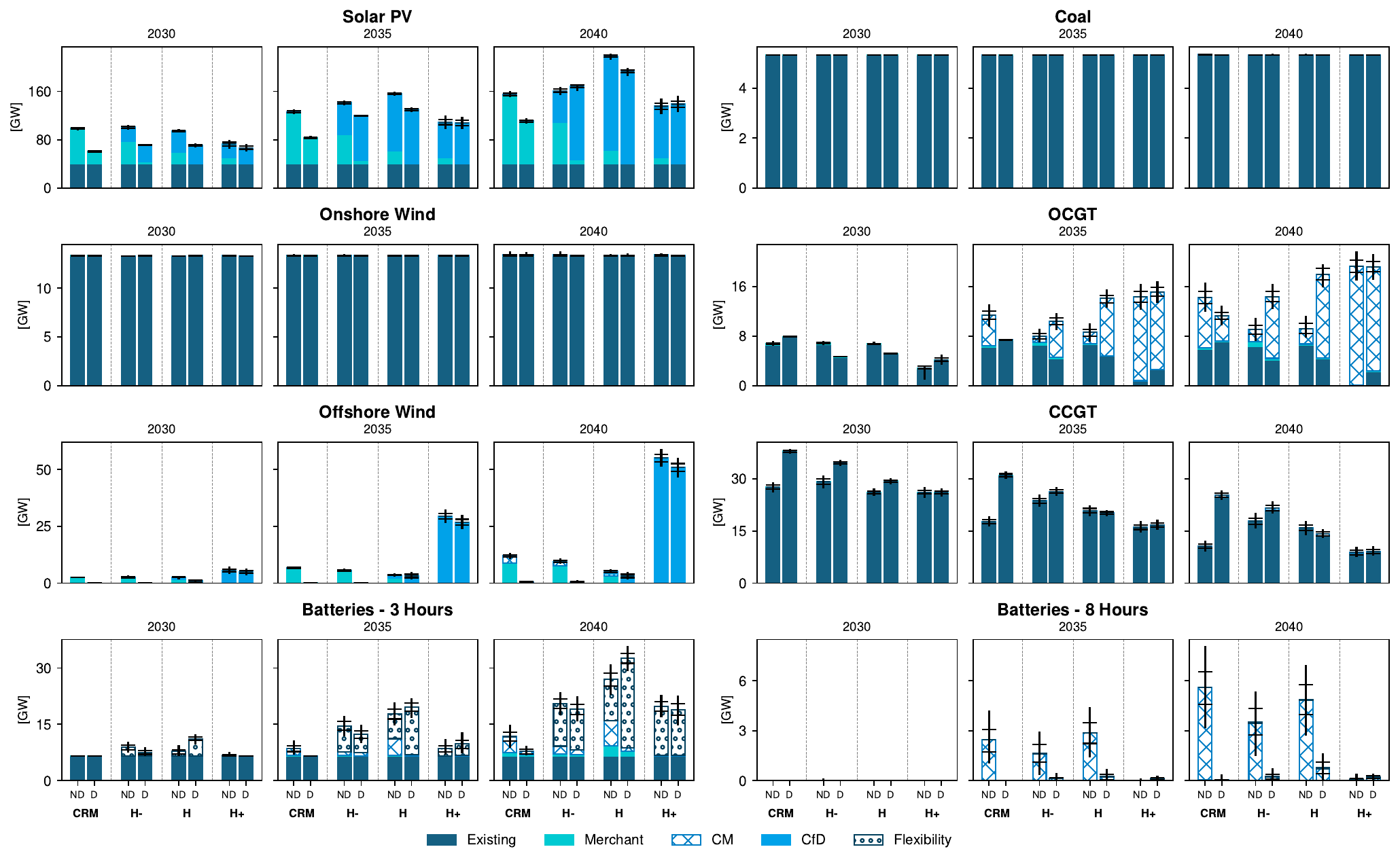}
\caption{\textbf{Installed capacity of generation and storage technologies in 2030, 2035, and 2040 under \textbf{[ND]} and \textbf{[D]} scenarios.} Stacked bars indicate average installed capacities across runs. In each stacked bar, the horizontal markers display the 25th and 75th percentiles, and the vertical markers display the 5th and 95th percentiles. Different hatching patterns highlight the mechanisms agents use to enter the market.}
\label{fig: Capacities evolution ND - D}
\end{figure}

The differences in installed capacity between \textbf{[ND]} and \textbf{[D]} scenarios, shown in Figure \ref{fig: Capacities evolution ND - D}, underpin the cost and emission dynamics described above. In the \textbf{[ND]} scenario, the system gradually replaces existing CCGT capacity with RES, storage, and a significant OCGT fleet operating at low capacity factors. This trend, though modulated by the long-term regulatory mechanisms available, is consistent across market configurations. Under the \textbf{[D]} scenario, by contrast, merchant incentives for green investment are weakened and lead the system to rely more on fossil and existing assets. As the simulation progresses, RES and storage capacity are either sustained through long-term regulatory mechanisms where the market configuration enables it, or displaced by continued reliance on existing CCGT capacity and newly added OCGT units, prolonging the dependence on fossil fuels and their contribution to system costs. Prominently, \textbf{[H+]} departs from all other scenarios in its investment dynamics: more ambitious CfD auctions trigger an early shift to offshore wind at the expense of solar PV, thus reducing the need for storage. This trend accelerates the decommissioning of the existing fossil-fuel fleet but, in turn, increases reliance on OCGT capacity for system security, which enters the market through the CRM.

Complementing Figure~\ref{fig: Capacities evolution ND - D}, Table~\ref{table: Capacity scenarios ND - D} contextualizes the relative size of the long-term regulatory mechanisms that act as entry gateways to the system. The yearly volumes procured through CfD and flexibility auctions in our scenarios are broadly of the same order of magnitude as Italy's most recent auctions: the 2025 FER~X and MACSE rounds \cite{sergio_italys_2025,terna_terna_2025}, which together awarded on the order of 8~GW of solar PV, around 1~GW of wind, and roughly 1.5~GW of battery storage \footnote{FER~X (\textit{Fonti Energetiche Rinnovabili}) is the Italian support scheme for mature renewable technologies. MACSE \textit{(Meccanismo di Approvvigionamento di Capacità di Stoccaggio Elettrico)} is Italy's competitive auction scheme for utility-scale storage. Auction results are further referenced as part of the Appendix.}. The correspondence is closest in the \textbf{[H-]} and \textbf{[H]} scenarios for solar PV and storage, where both magnitude and technology coincide. By contrast, onshore wind is essentially absent from our scenarios. This compositional gap widens in the most ambitious \textbf{[H+]} scenario, where investment shifts away from solar toward offshore wind: solar CfD volumes fall relative to \textbf{[H]}, while offshore wind rises sharply due to its higher capacity factor compared to the onshore variant. Nonetheless, every hybrid configuration, including the least ambitious \textbf{[H-]}, implies a markedly stronger commitment than a one-off intervention, since it sustains auctions of that scale over time. This reliance on long-term mechanisms deepens further when the carbon price signal is suppressed: in the \textbf{[H-]} and \textbf{[H]} scenarios, auctioned volumes rise to replace the merchant investment that no longer materializes. In \textbf{[H+]}, where most capacity already enters through the auctions and is remunerated via the long-term mechanisms, this substitution effect is less pronounced.

\begin{table}[!hbt]
\centering
\resizebox{\textwidth}{!}{%
\begin{tabular}{clllllllll}
\hline
\multirow{4}{*}{\textbf{Technologies}} &
  \multicolumn{1}{c}{\multirow{4}{*}{\textbf{Mechanism}}} &
  \multicolumn{8}{c}{\multirow{2}{*}{\textbf{Scenarios}}} \\
                                        & \multicolumn{1}{c}{} & \multicolumn{8}{c}{}                                                \\ \cline{3-10} 
 &
  \multicolumn{1}{c}{} &
  \multicolumn{2}{c}{CRM} &
  \multicolumn{2}{c}{H-} &
  \multicolumn{2}{c}{H} &
  \multicolumn{2}{c}{H+} \\
                                        & \multicolumn{1}{c}{} & ND         & D          & ND   & D    & ND    & D     & ND   & D    \\ \hline
\multirow{2}{*}{\textbf{Solar PV}}      & Merchant             & 7.55       & 4.72       & 4.55 & 0.47 & 1.54  & 0.02  & 0.69 & 0.01 \\
                                        & CfD                  & \multicolumn{2}{c}{N/A} & 3.48 & 8.04 & 10.33 & 10.16 & 5.64 & 6.56 \\
\multirow{2}{*}{\textbf{Offshore wind}} & Merchant             & 0.58       & 0.02       & 0.51 & 0.02 & 0.21  & 0     & 0    & 0    \\
                                        & CfD                  & \multicolumn{2}{c}{N/A} & 0.03 & 0.01 & 0.08  & 0.21  & 3.67 & 3.39 \\
\textbf{OCGT}                           & CRM                  & 0.54       & 0.28       & 0.13 & 0.67 & 0.17  & 0.9   & 1.28 & 1.13 \\
\multirow{2}{*}{\textbf{Battery-3h}}    & CRM                  & 0.29       & 0.06       & 0.14 & 0.09 & 0.45  & 0.06  & 0.02 & 0.02 \\
                                        & Flexibility          & \multicolumn{2}{c}{N/A} & 0.75 & 0.72 & 0.73  & 1.58  & 0.86 & 0.8  \\
\textbf{Battery-8h}                     & CRM                  & 0.37       & 0          & 0.23 & 0.02 & 0.32  & 0.05  & 0    & 0.01 \\ \hline
\end{tabular}%
}
\caption{\textbf{Yearly installed capacity additions [GW] of representative technologies across market mechanisms.} Values are yearly averages of installed capacity for each regulatory mechanism over the complete 2025--2040 horizon. Technologies and mechanisms not reported show no notable variation in installed capacity.}
\label{table: Capacity scenarios ND - D}
\end{table}

\begin{figure}[hbt!]
\centering
\includegraphics[width=1.0\linewidth]{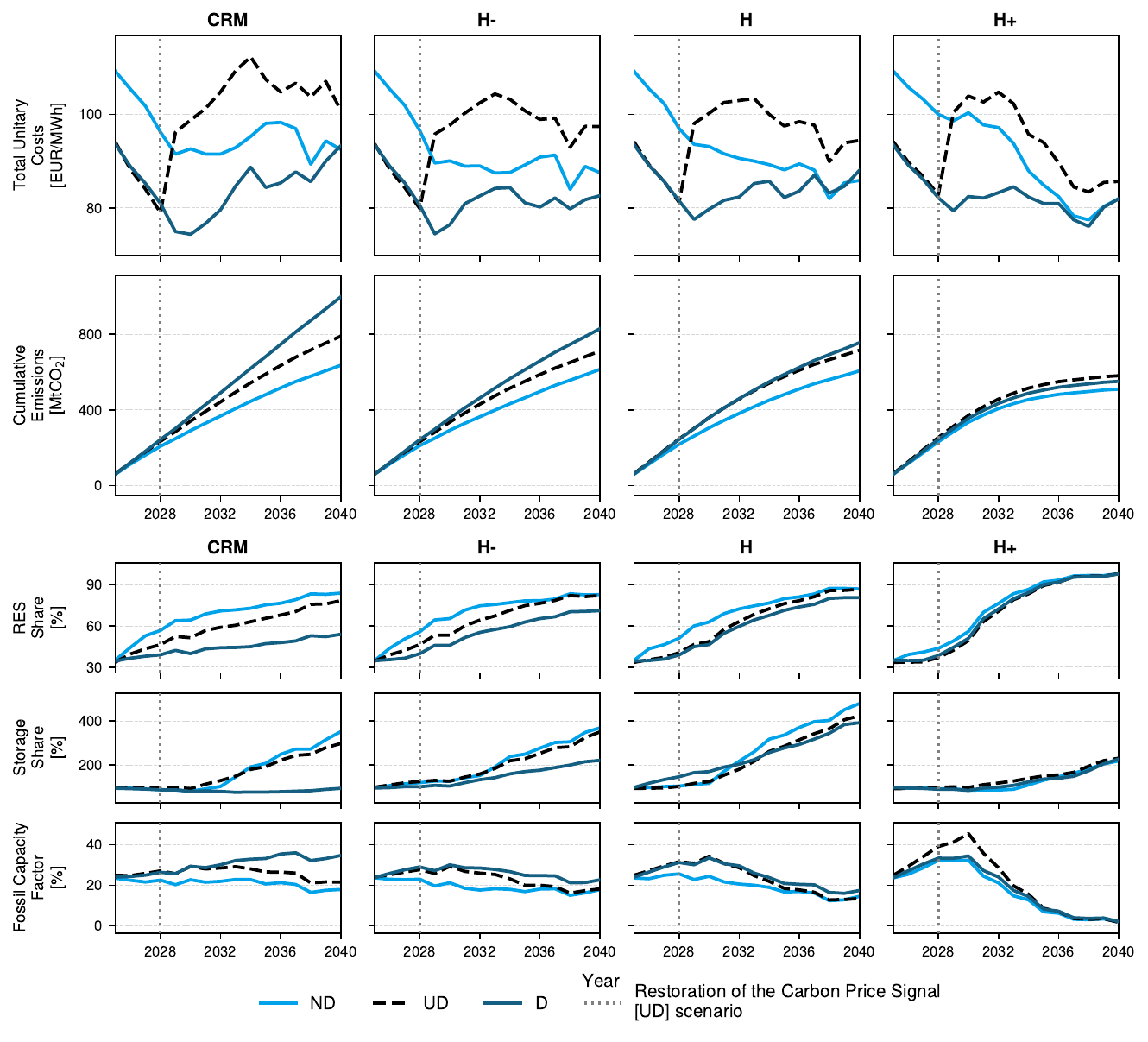}
\caption{\textbf{Evolution of system costs, cumulative CO$_2$ emissions, RES and storage shares of average demand, and fossil-fuel capacity factor across the [ND], [D], and [UD] scenarios.} The dashed line marks the reinstatement of the carbon price signal in the \textbf{[UD]} scenario.}
\label{fig: Evolution - All}
\end{figure}

The evolution of costs and emissions across all policy scenarios, as shown in the upper panel of Figure \ref{fig: Evolution - All}, reinforces the tendencies highlighted earlier. System costs in scenarios \textbf{[ND]} and \textbf{[D]} initially diverged, converging only in the last stages of the simulation. This behavior is partially explained by the sustained increase in emissions when the carbon price signal is suppressed, resulting in substantial ETS recovery costs. Introducing uncertainty about the duration of the \textit{Decreto Bollette}, as modeled in scenario \textbf{[UD]}, produces two distinct system outcomes. In terms of total unitary costs, the \textbf{[UD]} case exhibits the short-term price reduction observed in Figure \ref{fig: Prices and emissions ND - D} during the period when the policy measure is active. However, once the measure is removed, costs increase to levels equal to or above the \textbf{[ND]} baseline, reflecting both the restoration of the carbon price signal, as well as the delay in investments caused by the temporal application of the \textit{Decreto Bollete}. Correspondingly, cumulative emissions in the \textbf{[UD]} trajectory lie between those of \textbf{[ND]} and \textbf{[D]} scenarios, but once the carbon signal is restored, the pace of decarbonization accelerates to closely resemble that of the latter cases, although the emission offset remains.

The lower panel of Figure \ref{fig: Evolution - All} extends this analysis by highlighting key technological differences across scenarios. In the \textbf{[H+]} configuration, the transition toward low-carbon technology is minimally affected by the carbon price signal. In all other market designs, this decarbonization tendency breaks down to differing degrees under the influence of long-term regulatory mechanisms. As a result, fossil-fuel utilization is maintained or even increased when the carbon price signal is suppressed by the Decree (scenario \textbf{[D]}). Similarly, the \textbf{[UD]} scenario shows a storage trajectory that closely follows the \textbf{[D]} baseline, followed by a RES uptake that partially recovers once the carbon price signal is restored. 

\begin{figure}[hbt!]
\centering
\includegraphics[width=1.0\linewidth]{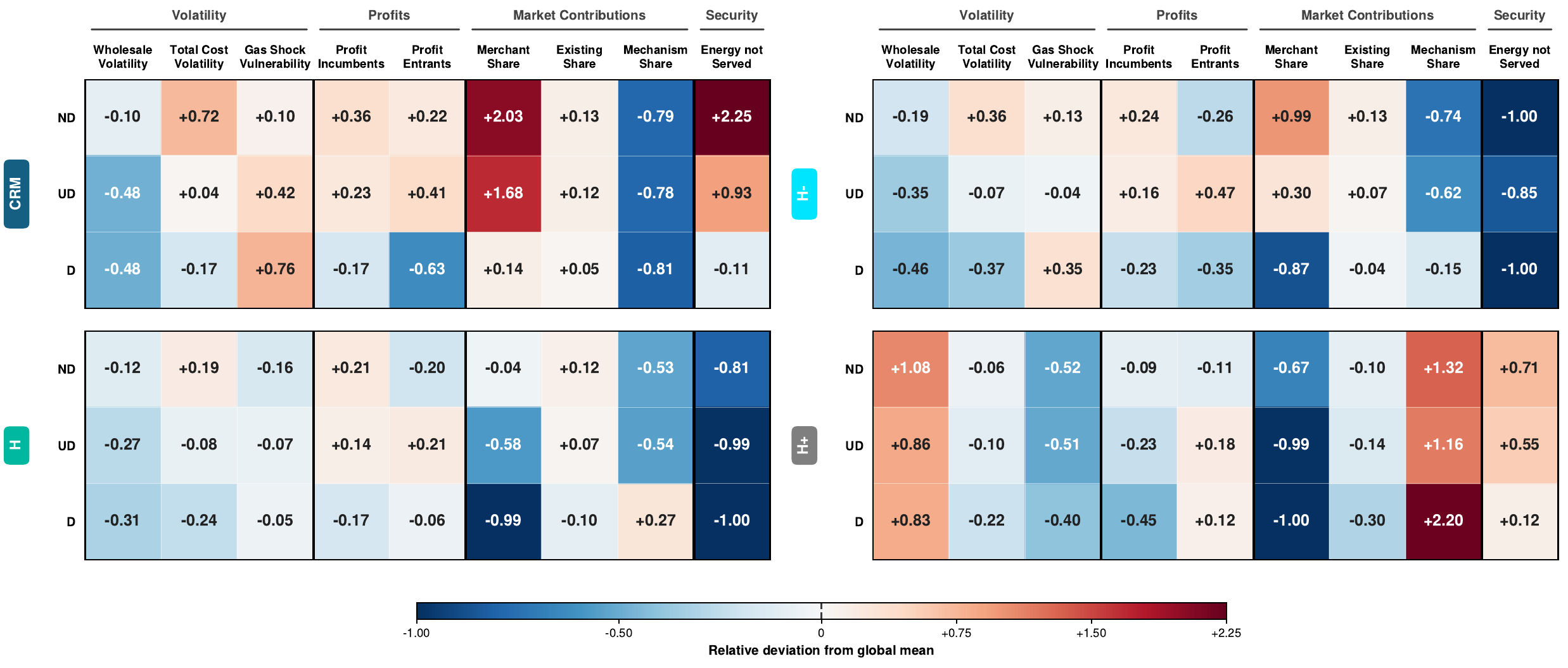}
\caption{\textbf{Heatmap of aggregate key metrics across the scenario matrix.} Metrics are normalized with respect to the average across all scenarios and capped at $-1.0$ and $+2.25$ for visualization. Wholesale and total price volatility are computed as the standard deviation of hourly prices, using normalized yearly values to discount inter-year variability. Shock vulnerability measures the relative cost between simulations with and without the natural gas price shocks. Profit metrics report the discounted net present value accruing to each agent category. Market contributions quantify the relative weight of merchant investments, existing assets, and long-term regulatory mechanisms in the final cost. Energy not served measures unmet demand across the simulations.}
\label{fig: Heatmap all}
\end{figure}

Figure \ref{fig: Heatmap all} summarizes additional system-performance metrics for the full scenario matrix, each normalized with the average across the scenario pool. The most consistent change associated with the Decree is a reduction in the merchant's investment share of total system costs. In configurations with green investment support, this shift increases dependence on long-term regulatory mechanisms to drive capacity expansion. In configurations without such support, such as the \textbf{[CRM]} case and to a lesser extent in the \textbf{[H-]} scenario, the same shift translates into higher utilization of existing fossil-based assets, increasing their market contribution and driving higher emissions, consistent with the trajectories shown in Figure \ref{fig: Evolution - All}. The higher share of variable technologies in the generation mix slightly increases wholesale price volatility, though total system cost volatility remains comparable across market configurations, reflecting the financial hedging provided by CfD contracts and the arbitrage role of ESS in the \textbf{[H+]} and \textbf{[H]} scenarios. With respect to agent-level outcomes, the reduction in the carbon price signal erodes the profits of incumbent agents on their existing assets and reduces the returns of new entrants, most of whom invest in renewable and storage assets. Finally, the absence of green investment support and the suppression of the carbon price signal make the system more vulnerable to natural gas price shocks, a condition that aligns with the increased utilization of fossil-fuel assets, as shown by the comparison in the corresponding metric between \textbf{[H+]} and \textbf{[CRM]} cases, but also between the \textbf{ND} and \textbf{[D]} cases within the same market scenarios. Adequacy varies across scenarios but remains within the capacity market's bounds throughout. Nominal values for all studied metrics are presented in the supplementary material. 

\begin{figure}[hbt!]
\centering
\includegraphics[width=1.0\linewidth]{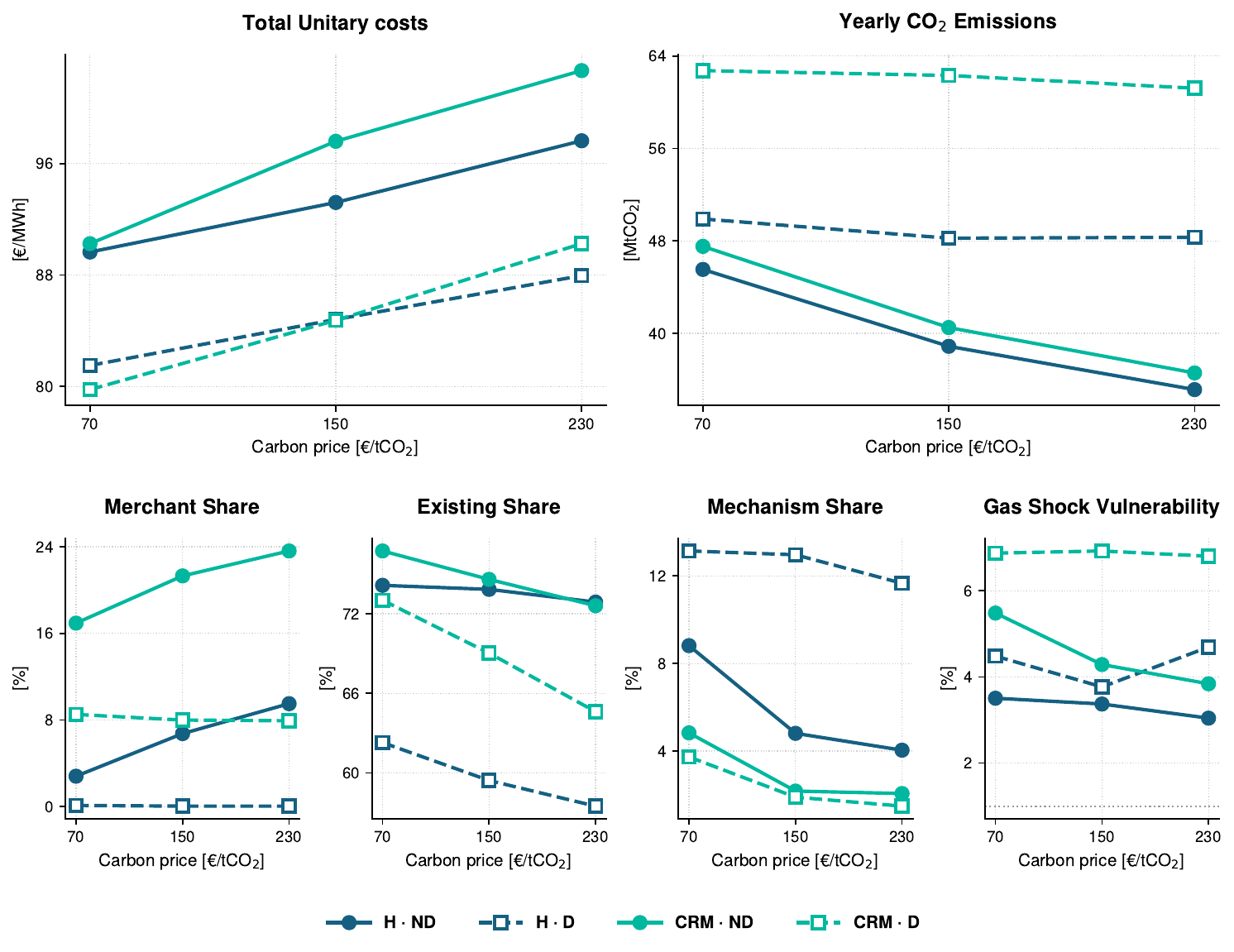}
\caption{\textbf{Aggregated system performance metrics under varying carbon price trajectories in the \textbf{[ND]} and \textbf{[D]} scenarios.} Results are organized by the 2040 carbon price on the x-axis. The upper panel shows total unitary costs and CO$_2$ emissions. The lower panels present: Shock vulnerability, measured as the relative cost between simulations with and without the natural gas price shock; Profit metrics, reporting the discounted net present value accruing to each agent category; and Market contributions, quantifying the relative weight of merchant investments, existing assets, and long-term regulatory mechanisms in the final cost.}
\label{fig: Carbon price scenarios}
\end{figure}

Finally, Figure \ref{fig: Carbon price scenarios} shows variations in key system metrics under the \textit{Decreto Bollette} across different carbon price trajectories for the \textbf{[H]} and \textbf{[CRM]} market configurations. Specifically, the baseline carbon price rises linearly from 70~\euro/tCO\textsubscript{2} in 2025, consistent with current EU-ETS levels, to 150~\euro/tCO\textsubscript{2} by 2040, while we complement it with two alternatives: a flat 70~\euro/tCO\textsubscript{2} held over the full horizon, and a steeper climb to 230~\euro/tCO\textsubscript{2} by 2040. In the \textbf{[ND]} scenarios, a higher carbon price produces lower CO$_2$ emissions with a modest increase in total system costs. The application of the Decree disrupts this relationship. When the carbon price is raised in a power system operating under carbon price suppression, the lack of prior investment in low-carbon technologies means that the higher ETS recovered costs fall disproportionately on a generation mix still dominated by fossil-fuel assets. The result is an increase in total system costs without a corresponding reduction in emissions, thereby inverting the standard carbon-pricing trade-off. This pattern is illustrated in the lower panel of Figure \ref{fig: Carbon price scenarios}, where configurations under the \textbf{[D]} scenario rely more on existing assets than their \textbf{[ND]} counterparts. Furthermore, in both \textbf{[H]} and \textbf{[CRM]} market configurations, the system's increased exposure to natural gas price shocks in the \textbf{[D]} scenario persists across all tested carbon prices.

\clearpage

\section*{DISCUSSION}
\label{Section - DISCUSSION}

This section expands on the three main outcomes of the study as identified in the previous section: (i) in the presence of a carbon price, and even with non-ambitious long-term regulatory mechanisms, the electricity system tends toward rapid decarbonization; (ii) the partial removal of the carbon price, as the one produced by the \textit{Decreto Bollette}, breaks this trend in most scenarios, trading substantial CO$_2$ emission increments for limited impacts on total system costs; and (iii) long-term regulatory mechanisms can partially shield the system from the most harmful effects of the carbon price suppression, but this condition requires immediate commitment and a shift in the market-design paradigm toward large-scale long-term regulatory mechanisms supporting green investments. We close the discussion by presenting the study's limitations, its implications for these conclusions, and future work.

\subsection*{Green investments and decarbonization trajectories}

A key trend across scenarios is that the current competitiveness of green investments, coupled with strong economic signals from carbon pricing and long-term regulatory mechanisms, provides a robust basis for decarbonization in systems that remain fossil-fuel-dependent. Although within the studied time horizon the system does not fully decarbonize, it does achieve rapid, cost-effective integration of renewable energy, bringing additional system-wide benefits, such as increased resilience to natural gas price shocks. These results emerge in a setting that deliberately assumes an aggressive demand-growth scenario, constraining the system to expand its generation assets to meet rising demand, a condition achieved without substantial incremental costs.

More specifically, results depict a power system that relies heavily on solar PV and storage to support decarbonization scenarios, which, in turn, drive the aggressive decommissioning of existing CCGT plants and their partial replacement with OCGT capacity. Onshore wind, given its relatively low capacity factor due to spatial aggregation in both the model and the resource information from \citet{antonini_weather-_2024}, is only moderately competitive. Offshore wind is only substantially competitive when aggressive, long-term regulatory mechanisms are in place, substituting solar PV in the \textbf{[H+]} scenario. The reshaping of the thermal fleet stems from three factors. First, the relatively low capacity factor required from fossil-fuel plants under large RES penetration. Second, the CRM design provides differentiated treatment to existing and new assets, favoring the latter and thereby opening the door to inefficient retrofitting from a system perspective but economically grounded from the agent's perspective. Third, as a modeling caveat, the absence of large-scale storage investment options leaves peaking plants as the available flexibility choice.

These findings, aligned with the literature analyzing future energy systems \cite{rodrigues_2040_2026,irena_world_2023}, reinforce the need to enable systems and markets oriented toward rapid decarbonization and to make them resilient not only to energy price shocks but also to policy designs that may have unintended effects and slow down the transition. Rapid decarbonization is possible, but it requires a set of necessary conditions to accelerate and sustain it. The following two sections discuss the role of the key enabling factors: a strong carbon price signal and long-term regulatory mechanisms to support green investment.

\subsection*{Carbon pricing suppression}
The suppression of the carbon price signal in the electricity market produces short- and long-term effects. The former, perhaps more prevalent in discussions of price shocks and aligned with the results from analyses of the \textit{Iberian exception} \cite{haro_ruiz_effects_2024,fabra_winners_2025}, is the marked short-term cost reduction. This cost decrease, stemming from reductions in inframarginal rents and reduced profits for incumbent agents, could be directly reflected in the costs faced by final users and therefore carries significant weight in political discussions.

More notable than the overall short-term cost reduction is the effect on the wholesale price signal. Generators face a substantial price reduction with significant implications for investment profitability. This impact is more pronounced for merchant investments, making the system more reliant on long-term regulatory mechanisms. In terms of technologies, the suppression of the carbon price substantially reduces solar and storage investments, both of which are key technologies for the transition and for energy security, making the system more dependent on existing assets, most of which are gas-fired power plants.

Taken together, these compound effects push the system toward greater dependence on fossil-fuel resources. As a result, system emissions increase substantially over the long term, nullifying the earlier short-term cost benefits, since EU-ETS costs must be eventually repaid by final users. This also makes system costs more closely correlated with gas price shocks, an expected pattern as natural gas assets become more dominant. These long-term price pressures are also present in scenarios with a temporal, but unknown, duration of the \textit{Decreto Bollette}, leading agents to delay investments and resulting in comparatively higher costs. Although renewable investment recovers after the carbon price is fully restored, the risks faced by agents and the impacts on prices and emissions have already materialized.

This analysis yields a strong insight into the suppression of the carbon price: \textit{Decreto Bollette}-like policies offer short-term price reductions at the cost of long-term damage to the investment signal, leading to subsequent rising prices and substantial increases in CO$_2$ emissions. Nonetheless, the negative outcomes from carbon price suppression can be partially hedged by the presence of long-term regulatory mechanisms, as discussed next.

\subsection*{The role of long-term regulatory mechanisms}

The system behavior in the presence of a full carbon price and long-term regulatory mechanisms is worth understanding first. Assuming the mechanisms are in place throughout the simulation, market participants learn, relative to the \textbf{[CRM]} scenario, to wait for the long-term mechanisms to remunerate green investment rather than relying solely on wholesale revenues. This creates a self-reinforcing dynamic: once investors anticipate that capacity will be procured through auctions, merchant entry diminishes, which, in turn, makes auctions the binding channel for new capacity, further strengthening the incentive to wait for them. The evolution of the generation mix thus becomes increasingly contingent on the continued presence and sizing of the mechanisms. Finally, while our scenarios are not directly comparable with real auction outcomes, since the mechanisms are sustained across the whole horizon, the \textbf{[H-]} and \textbf{[H]} configurations procure average volumes broadly in line with Italy's recent auctions, whereas \textbf{[H+]} raises the level of commitment substantially, procuring large volumes of offshore wind whose higher capacity factor increases the overall share of renewable energy in the system.

The comparison across hybrid market scenarios showcases the strength of long-term mechanisms in shielding investment signals and incentives from the wholesale market when the carbon price is suppressed. This holds true even when all mechanisms expose assets to wholesale short-term prices, a key feature desired in current market design. Although scenarios \textbf{[H-]} and \textbf{[H]} are subject to the diminishing investment incentives for green assets discussed in the previous section, the magnitude is substantially lower than in the \textbf{[CRM]} scenario, which includes no direct support for RES and ESS. Notably, the \textbf{[H+]} scenario shows almost no increase in emissions when the carbon price signal is suppressed, and it sustains, over time, the total cost reductions observed in the short term.

As such, ambitious long-term regulatory mechanisms could push the system towards decarbonization while also shielding it from the negative impacts of suppressing the carbon price signal. However, this comes with caveats. As shown in Figure \ref{fig: Prices and emissions ND - D}, system costs shift almost entirely toward the long-term mechanisms, especially CfD contracts. Although this does not entail a direct cost increase for final users, the scale of the mechanisms requires a conceptual shift in market design, one in which the short-term wholesale price signal becomes largely irrelevant for remunerating utility-scale assets. Such transformation comes along with a level of commitment from regulators and policymakers to assume the role of active planners, in line with arguments for deeper hybrid market designs \cite{joskow_hierarchies_2022}. The contrast across the tested Hybrid scenarios highlights the need for a large level of commitment. Even in the least ambitious \textbf{[H-]} case, the auction volumes are substantial, comparable in scale to Italy's 2025 FER~X and MACSE rounds, but sustained year after year across the 2025-2040 horizon rather than procured once. In addition, the response from the auction mechanism is dynamic: as the carbon price signal is suppressed, the mechanisms react, in speed and size, to compensate for the merchant investment that no longer materializes. The shielding effect is therefore obtained through a durable, escalating, and actively planned commitment. Precisely in that regard, our assessment represents the best-case scenario for these mechanisms: we assume they exist and persist throughout the simulation under known conditions, so agents can reliably anticipate them and invest accordingly. In practice, the value of long-term mechanisms depends on regulatory commitment, and the \textbf{[UD]} scenario, where the timing of carbon-signal restoration is uncertain, shows that the cost of this uncertainty is far from negligible.

In this context, where large-scale commitments towards long-term support for green investment are needed for the transition and can also hedge against a suppressed carbon price, a deeper tension emerges from the political economy of these policies. The \textbf{[H+]} scenario can be tentatively considered as an ideal outcome that requires a regulatory posture in direct contradiction with the logic of the \textit{Decreto Bollette} itself. Ambitious long-term support for green investment could be understood as an indirect subsidy to technologies that, as our analysis shows, do not deliver strong system-level benefits when carbon pricing is suppressed. These are the same technologies that, conversely, become the cornerstone of the transition when long-term mechanisms are strong. The same political conditions that motivate cutting the carbon price signal are therefore unlikely to coexist with the level of commitment that \textbf{[H+]} demands. A policymaker willing to suppress the carbon price signal, but refraining from long-term green investment support at the level required, would lead the system toward a condition more aligned with the \textbf{[H-]} and \textbf{[CRM]} scenarios, where the vicious cycle of fossil-fuel dependence, long-term vulnerability to shocks, and rising emissions emerges.

This political tension is reinforced by the symmetric role of uncertainty. We assume the persistence of long-term mechanisms across the horizon, but the \textbf{[UD]} scenarios show that uncertainty about the persistence of the carbon price suppression already shapes agent behavior and erodes the price signal even before the policy materializes. By the same logic, uncertainty over the persistence of long-term mechanisms would erode the very investment incentives they are designed to provide, weakening the shielding effect that makes \textbf{[H+]} attractive in the first place. The credibility of the commitment, and not only its initial ambition, is therefore a necessary condition for the favorable outcomes observed in the Hybrid scenarios.

\subsection*{Limitations and future work}

From an electricity system perspective, MARLEY relies on three simplifications: a copper-plate network that ignores congestion and locational signals, the absence of technical security constraints, and the lack of an interconnection module for cross-border exchange. Each of these bears on the value of the assets we study, including, but not limited to, siting incentives and capture rates for renewables and storage, as well as penetration constraints. In particular, the lack of an interconnection module may lead us to underestimate both the incentive for increased thermal production induced by the carbon price suppression of the \textit{Decreto Bollette}, and its resulting impact on CO$_2$ emissions, as similar analyses of the \textit{Iberian Exception} have shown \citep{linares_assessment_2023,fabra_winners_2025,hidalgo-perez_iberian_2024}.

Moreover, we do not model the demand side or the PPA market. Although we represent substantial ambition in the CfD mechanisms, the PPA market could provide an alternative for these types of investments without direct government intervention. Modeling representative consumers and the interaction between demand response, electricity prices, and volatility may prove insightful for future analyses of the carbon price signal and price formation in electricity systems, as discussed in \citet{geis_price_2026}. 

We also assume static conditions in the EU-ETS market across scenarios. This could have important repercussions if similar measures become widespread, as the emissions impact is substantial in most scenarios. Plausibly, the feedback effects, if emission targets are maintained, would increase the cost of the \textbf{[D]} scenarios modeled, as the carbon price trajectory would rise with respect to the baseline. However, further work is needed to validate this hypothesis.

From a methodological perspective, the results depend on the MARL setup's learning dynamics. The use of IPPO represents one algorithmic choice among several available in the competitive MARL literature, and the design of agent objectives, action spaces, and exploration parameters shapes the equilibria to which agents converge. While the consistency of the qualitative trends across tests and scenarios provides some reassurance regarding the robustness of the findings, sensitivity to alternative algorithmic choices, reward formulations, and richer representations of investor heterogeneity warrants further exploration. 

Furthermore, we refrain from explicitly incorporating risk aversion into the MARL algorithms. Although single-agent risk-averse implementations exist \cite{bisi_risk-averse_2022}, their implications for multi-agent competitive environments deserve further attention. We anticipate that risk aversion would materially affect these results, particularly under the policy, fuel-source, cost, and market uncertainties we examine, by strengthening agents' preference for the revenue certainty of long-term support over merchant exposure.

Finally, we refrain from discussing the technical issues related to the legality and applicability of the Italian Decree within the European legal framework, as this lies outside the scope of our assessment. We instead treat this as a general measure, applicable to other power systems, in which carbon price signals are not fully transmitted into electricity dispatch.

\clearpage

\section*{METHODS}
\label{Section - METHODS}

This section builds on the general overview presented in Figure \ref{fig: General Overview} for the MARLEY framework, introducing the multi-agent electricity market model, the MARL implementation, the policy scenarios, and the experimental setup.  

\subsection*{The long-term MARL Electricity Market Model}
\label{Subsection - Long-term Electricity Market Environment}

MARLEY is built around a MARL environment, implemented using the Gymnasium and RLlib libraries \cite{towers_gymnasium_2023,liang_rllib_2018}, targeted towards mid- and long-term decarbonization analysis in wholesale electricity markets, focusing on utility-scale investment decisions and market mechanisms designed to promote them.

Stylized short-term markets are the core of MARLEY's power system representation. They are constructed using 24-hour representative periods obtained by aggregating publicly available time series for the Italian system \cite{antonini_weather-_2024,di_bella_mitigation_2025} with standard techniques \cite{hoffmann_typical_2021}, and aim to represent a bimonthly period. Each representative day is solved through a cost-minimizing optimization in which generation assets bid quantities and prices reflecting their availability and marginal costs (including any carbon price). Short- and mid-term storage assets, in contrast, operate freely within their technical constraints at zero marginal cost, with GENCOs setting the desired levels of pumped-hydro assets for the following representative period. The dispatch is subject to demand and flexibility balance constraints, whose shadow prices define the energy and flexibility prices remunerated to contributing assets, and adopts a copper-plate assumption that abstracts from transmission limits on energy injection and resource integration.

In this market, GENCOs maximize economic profit by owning and operating a portfolio of generation and storage assets under a decentralized market structure. Their portfolios span the key power technologies \textit{(utility-scale solar PV, onshore wind, offshore wind, coal, combined-cycle gas turbines (CCGT), open-cycle gas turbines (OCGT), and 3- and 8-hour batteries)}. In addition, each incumbent GENCO holds a set of existing assets, including pumped hydro, and may actively decommission CCGT and OCGT units over the horizon.

In addition, a high-level policymaker implements targeted interventions to minimize system costs and reduce CO$_2$ emissions, using the carbon price to unify these objectives into a single function. Precisely, the policymaker, with an annual resolution, sets the targets for both the long-term regulatory mechanisms promoting RES investments, intended to represent production-based CfD auctions \cite{european_university_institute_robert_schuman_centre_for_advanced_studies_contracts-for-difference_2024}, and a long-term flexibility-service procurement mechanism without dispatch restrictions, as classified by \citep{mastropietro_taxonomy_2024}. Storage is remunerated through a premium per MWh of installed energy capacity while preserving partial exposure to short-term market signals. Complementing these mechanisms, the environment integrates a capacity market based on a Reliability Option Design \cite{cramton_capacity_2013}, in which a fixed adequacy target drives investment in firm capacity.

In the long term, GENCOs invest in generation and storage assets through four mutually exclusive channels: 
\begin{itemize}
    \item \textbf{Merchant investments} carried out outside any long-term mechanism and remunerated solely through wholesale inframarginal and scarcity rents;
    \item \textbf{RES investments supported via CfD auctions} where the GENCO commits their production to a double-sided contract financially settled against the auction strike price and the short-term market price;
    \item \textbf{Capacity market investments}, where resources receive a premium based on their endogenous capacity credit under a reliability option, forgoing scarcity rents in exchange. Existing assets also participate, but receive a premium equal to half that of entrants, and only in years when an adequacy gap is identified; and
    \item \textbf{Short-term storage investments supported by flexibility auctions} where resources receive a premium for the energy storage capacity for participating in the wholesale market and covering the system's flexibility needs. 
\end{itemize}

No forced investment or alternative driving signal beyond economic profit is modeled in MARLEY. As a result, training converges to market conditions where agents earn consistently positive profits, rather than to the near-zero-profit condition expected under perfect competition. This outcome reflects several structural features of the framework: the investment formulation itself, where inaction yields a risk-free zero-profit outcome, the limited number of agents, the predictability and implicit coordination that emerge from repeated training under identical system conditions, the non-linearities embedded in market and mechanism design, and the imperfections inherent to current MARL training techniques.

\subsection*{The Multi-Agent Reinforcement Learning Implementation}
\label{Subsection - The Multi-Agent Reinforcement Learning Implementation}

The market environment is formalized as a Partially Observable Stochastic Game (POSG), in which agents observe a local state comprising their portfolio composition, market signals, and exogenous system conditions, and select an action that combines discrete investment decisions, relying heavily on action-masking to reduce the decision space \cite{huang_closer_2022}. Specifically, GENCOs receive a private reward corresponding to its net economic profit from market participation. The policymaker is modeled analogously, with its own observation, action, and reward structures tied to the system-level cost and emission objectives. 

In line with \citet{gonzalez-ruiz_assessing_2026}, we solve this POSG using Independent Proximal Policy Optimization (IPPO), and adapting the RLlib and Gymnasium implementations \cite{liang_rllib_2018,towers_gymnasium_2023}. PPO, introduced in \citet{schulman_proximal_2017} and extended to multi-agent settings in \citet{yu_surprising_2022}, is an on-policy Actor-Critic algorithm that trains a stochastic policy through two neural networks. The Critic provides a baseline for reward estimation, while the Actor selects actions that maximize reward given the observed state. Training relies on the standard clipped surrogate objective, with entropy regularization to encourage exploration and a value-function loss for the Critic \cite{schulman_proximal_2017,bick_towards_2021}. In its multi-agent independent learning variant, this training process is replicated across all agents, with each agent treating the others as part of the environment, without parameter sharing or a centralized critic. 

We adopt IPPO as a scalable solution that robustly converges to an equilibrium of the market problem under reasonable computational constraints, despite the well-known challenges associated with independent learning \textit{(credit assignment, equilibrium multiplicity, and the inherent non-stationarity of the learning process)} \cite{albrecht_multi-agent_2024}. 

The Supplementary Material reports a robustness analysis of the methods and a comparison with the centralized-training, decentralized-execution MAPPO variant of \citet{yu_surprising_2022}, confirming both IPPO's suitability in this setting and the reliability of our main results and conclusions. 

\subsection*{Scenarios}

For the scenarios, we adopt a 2025–2040 horizon, long enough to capture both the short-term suppression of the carbon price signal and its longer-term consequences for investment, as agents respond to the resulting signals through decommissioning decisions and reliance on the long-term regulatory mechanisms.

We design a scenario matrix that isolates the effects of partially removing the carbon price signal under different ambition levels for renewable and storage support. The two central scenarios fix the policy treatment of the carbon price: \textbf{[ND]} retains the full carbon price signal, while \textbf{[D]} applies the partial suppression of the carbon price signal introduced by the \textit{Decreto Bollette} for the whole simulation horizon. 

In all market configuration scenarios, and as the sole mechanism in the \textbf{[CRM]} case, a capacity market is implemented with a fixed adequacy target. From that baseline, long-term ambition varies across the hybrid scenarios \textbf{[H+]}, \textbf{[H]}, and \textbf{[H-]}, which differ in the parameters governing the CfD and flexibility mechanisms available to the stylized policymaker. We emphasize that these parameters define upper bounds rather than active constraints: they set the maximum target and the maximum yearly target increment the policymaker may pursue, but the realized auction volumes are outcomes of the reinforcement-learning process rather than fixed inputs. Within these limits, the policymaker learns how aggressively to procure capacity, while the volume actually awarded depends on both its decisions and the GENCOs' willingness to invest. A target is therefore not necessarily reached: under the 120\% ceiling, for example, RES penetration may fall short if market conditions do not support the corresponding investment, even when the policymaker sets the target and opens the auctions to procure it. As before, we assume the long-term mechanisms are present throughout the entire simulation horizon. The complete configuration across scenarios is reported in Table~\ref{table: Market configurations}.

\begin{table}[hbt!]
\centering
\begin{tabular}{@{}clcccc@{}}
\toprule
\multirow{2}{*}{\textbf{\begin{tabular}[c]{@{}c@{}}Market\\ mechanism\end{tabular}}} &
  \multicolumn{1}{c}{\multirow{2}{*}{\textbf{Market feature}}} &
  \multicolumn{4}{c}{\textbf{Scenarios}} \\
 &
  \multicolumn{1}{c}{} &
  \multicolumn{1}{l}{\textbf{CRM}} &
  \multicolumn{1}{l}{\textbf{H-}} &
  \multicolumn{1}{l}{\textbf{H}} &
  \multicolumn{1}{l}{\textbf{H+}} \\ \midrule
\multirow{6}{*}{\textbf{\begin{tabular}[c]{@{}c@{}}Capacity\\ Market\end{tabular}}} &
  \begin{tabular}[c]{@{}l@{}}Adequacy Target:\\ Energy not served ratio\\ over the mean demand\end{tabular} &
  \multicolumn{4}{c}{\multirow{2}{*}{0.0008}} \\
 &
  \textit{{[}\%{]}} &
  \multicolumn{4}{c}{} \\
 &
  \begin{tabular}[c]{@{}l@{}}Price-strike for\\ Reliability Option\end{tabular} &
  \multicolumn{4}{c}{\multirow{2}{*}{300}} \\
 &
  \textit{{[}EUR/MWh{]}} &
  \multicolumn{4}{c}{} \\
 &
  \begin{tabular}[c]{@{}l@{}}Price-cap auction:\\ Premium for\\ Reliability Option\end{tabular} &
  \multicolumn{4}{c}{\multirow{2}{*}{20}} \\
 &
  \textit{{[}EUR/MWh-firm{]}} &
  \multicolumn{4}{c}{} \\ \midrule
\multirow{6}{*}{\textbf{\begin{tabular}[c]{@{}c@{}}Contract for \\ Difference\\ Market\end{tabular}}} &
  \begin{tabular}[c]{@{}l@{}}Maximum Target:\\ RES energy production\\ over the mean demand\end{tabular} &
  \multirow{2}{*}{N/A} &
  \multirow{2}{*}{60} &
  \multirow{2}{*}{80} &
  \multirow{2}{*}{120} \\
 &
  {[}\%{]} &
   &
   &
   &
   \\
 &
  \begin{tabular}[c]{@{}l@{}}Maximum yearly target growth:\\ RES energy production\\ over the mean demand\end{tabular} &
  \multirow{2}{*}{N/A} &
  \multirow{2}{*}{4} &
  \multirow{2}{*}{6} &
  \multirow{2}{*}{8} \\
 &
  {[}\%{]} &
   &
   &
   &
   \\
 &
  \begin{tabular}[c]{@{}l@{}}Price-cap auction:\\ Strike for the\\ two-way CfD\end{tabular} &
  \multirow{2}{*}{N/A} &
  \multicolumn{3}{c}{\multirow{2}{*}{150}} \\
 &
  \textit{{[}EUR/MWh{]}} &
   &
  \multicolumn{3}{c}{} \\ \midrule
\multirow{6}{*}{\textbf{\begin{tabular}[c]{@{}c@{}}Flexibility\\ Market\end{tabular}}} &
  \begin{tabular}[c]{@{}l@{}}Maximum yearly target growth:\\ ESS energy capacity\\ over the mean demand\end{tabular} &
  \multirow{2}{*}{N/A} &
  \multirow{2}{*}{60} &
  \multirow{2}{*}{80} &
  \multirow{2}{*}{120} \\
 &
  {[}\%{]} &
   &
   &
   &
   \\
 &
  \begin{tabular}[c]{@{}l@{}}Maximum yearly target growth:\\ RES energy production\\ over the mean demand\end{tabular} &
  \multirow{2}{*}{N/A} &
  \multirow{2}{*}{4} &
  \multirow{2}{*}{6} &
  \multirow{2}{*}{8} \\
 &
  {[}\%{]} &
   &
   &
   &
   \\
 &
  \begin{tabular}[c]{@{}l@{}}Price-cap auction:\\ Premium for energy\\ storage capacity\end{tabular} &
  \multirow{2}{*}{N/A} &
  \multicolumn{3}{c}{\multirow{2}{*}{30,000}} \\
 &
  \textit{{[}EUR/MWh/year{]}} &
   &
  \multicolumn{3}{c}{} \\ \bottomrule
\end{tabular}%
\caption{\textbf{Design parameters of the different regulatory mechanisms across policy scenarios.} The table includes the parameters for the products for each specific mechanism, the parameters used by the policymaker to set the penetration targets, and the price caps of the corresponding auctions.}
\label{table: Market configurations}
\end{table}

The \textbf{[UD]} scenarios are designed to capture the temporary nature of the carbon price suppression. During training, agents are exposed to four predefined cases in which the suppression is reversed in simulation year 4, 8, or 12, or never reinstated, preventing them from anticipating the exact moment of restoration. The Results section focuses on the policy-relevant outcome of reinstatement in year 4, representing a government that uses suppression as a short-term consumer price relief measure before restoring the full carbon price signal. This case illustrates how short-lived relief produces only transient effects on the long-term trajectory once agents internalize this uncertainty

In all of the above scenarios, the carbon price starts at 70~\euro/tCO\textsubscript{2} in 2025, in line with the current EU-ETS price, and rises linearly to 150~\euro/tCO\textsubscript{2} by 2040. Building on this baseline, we use the \textbf{[H]} and \textbf{[CRM]} scenarios, which respectively include and exclude long-term mechanisms supporting green investment, to test two additional carbon price trajectories. In the first, the carbon price remains flat at 70~\euro/tCO\textsubscript{2} over the entire horizon; in the second, it rises to 230~\euro/tCO\textsubscript{2} by 2040, intensifying the decarbonization pressure on the policymaker.

Across all scenarios, we calibrate system conditions to emulate 2025 as the starting point \cite{terna_terna_2026, ember_ember_2026}. Demand reflects 2025 values, assuming net imports remain constant across the analyzed horizon. We further assume a 2\% annual demand growth rate, departing from the recent stagnation and aligning with ambitious electrification pathways for the Italian system \cite{di_bella_multi-objective_2021}. Moreover, we adopt the declining-cost trajectories for renewable and storage technologies reported in the PyPSA database \cite{brown_pypsa-eur_2024}, and we augment the fuel cost assumptions from the same source with a regime-switching shock module that becomes active after 2030. This triggers gas price shocks roughly once every 12 years, with typical peaks at twice the long-term price and a duration of 2 years. This component exposes the system, uniformly across scenarios, to stylized disruptions of the kind observed during the 2022 and 2025-2026 European gas crises. For the gas shock metric in Figure \ref{fig: Heatmap all}, we compare results with and without the previous module active, highlighting the system's vulnerability to gas price variations. 

Furthermore, we model the active decommissioning of CCGT and OCGT plants, while the existing coal fleet remains nominally in operation for the duration of the analysis, but is effectively mothballed under prevailing fuel and carbon price conditions. Finally, we model the Italian system with 16 agents, divided equally between entrants and incumbents. All agents operate their full portfolios, enabling them to pursue complementary investment strategies across all available asset classes. Further details and scenario descriptions are provided in the supplementary material.

\subsection*{Experiments}

To preserve scenario-specific dynamics, we run an independent training session for each scenario, using a conservative configuration that prioritizes learning stability over throughput \cite{yu_surprising_2022,gonzalez-ruiz_assessing_2026}. We fix a common computational budget across experiments, ensuring sufficient learning iterations for agents to settle into stable strategies, and deliberately avoid learning-rate or entropy schedulers that could force premature convergence.

\begin{figure}[hbt!]
\centering
\includegraphics[width=1.0\linewidth]{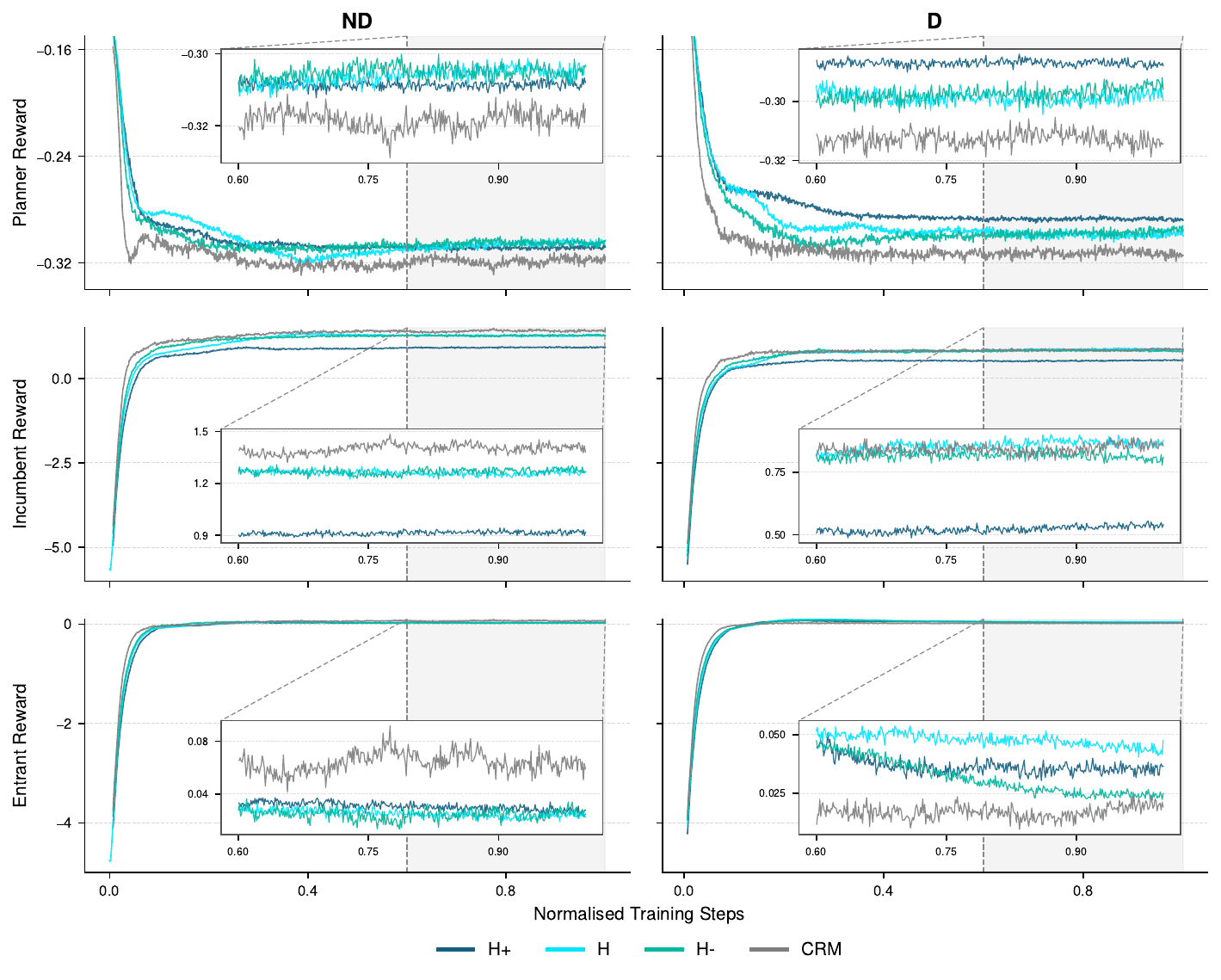} 
\caption{\textbf{Evolution of mean agent reward by category during training across the main scenarios.} Training steps are normalized by the maximum value reached within the computational budget. Agents are grouped into the policymaker, incumbent GENCOs, and entrant GENCOs. The inset highlights agent behavior in the final stages of training.}
\label{fig: Training all}
\end{figure}

Figure \ref{fig: Training all} illustrates the resulting training dynamics, showing the evolution of rewards by agent category across the main scenarios. This training dynamics can be divided into two distinct stages. In the first, GENCOs over-invest in the system, producing negative economic performance but lower system costs and emissions. From this starting point, GENCOs progressively filter out inefficient strategies by scaling back investments. Depending on the market design, this contraction of merchant investment triggers long-term mechanisms, a regime change that policies need to internalize during training. By the end of the training session, GENCOs rewards stabilize at positive values, indicating that investments achieve an internal rate of return above the discount rate, as evidenced by the aggregated rewards of entrant agents. Nonetheless, reward levels still vary across training steps, underscoring the challenges of convergence in multi-agent environments. Hyperparameters, network configurations, and robustness analyses are reported in the supplementary material.

Once the training session is complete for each scenario, market and system outcomes are obtained from 1{,}000 independent environment trajectories, allowing us to capture both the stochasticity of the environment and that of the agent strategies.


\clearpage

\section*{RESOURCE AVAILABILITY}


\subsection*{Lead contact}


Requests for further information and resources should be directed to and fulfilled by the lead contact, Javier Gonzalez-Ruiz (javier.gonzalez@cmcc.it), and/or Carlos Rodriguez-Pardo (carlos.rodriguezpardo.jimenez@gmail.com). 

\subsection*{Materials availability}

This study did not generate new materials.

\subsection*{Data and code availability}


\begin{itemize}
    \item All original code will be stored in a publicly available repository upon acceptance.
    \item Any additional information required to reanalyze the data reported in this paper is available from the lead contact upon request.    
\end{itemize}

\section*{ACKNOWLEDGMENTS}


Javier Gonzalez-Ruiz, Carlos Rodriguez-Pardo, and Massimo Tavoni acknowledge support from the European Research Council, ERC grant agreement number 101044703 (EUNICE) CUP D87G22000340006. Alice Di Bella acknowledges funding from European Union PNRR - Missione 4–Componente 2–Avviso 341 del 15/03/2022 - Next Generation EU, in the framework of the project GRINS - Growing Resilient, INclusive and Sustainable project (GRINS PE00000018 – CUP C83C22000890001). Paolo Mastropietro acknowledges funding from the ONESYSTEM research project (grant CPP2022-009711), funded by MICIU/AEI /10.13039/501100011033 and by the European Union NextGenerationEU/PRTR.

\section*{AUTHOR CONTRIBUTIONS}

Conceptualization, J.G., P.M., and M.T.; methodology, J.G. C.R. A.D.B, and P.M.; investigation, J.G., and P.M.; writing-–original draft, J.G.; writing-–review \& editing, A.D.B, C.R., P.M., J.P.C., and M.T.; funding acquisition, P.M., J.P.C., and M.T.; supervision, P.M., J.P.C., and M.T.
\section*{DECLARATION OF INTERESTS}


The authors declare no competing interests.

\section*{DECLARATION OF GENERATIVE AI AND AI-ASSISTED TECHNOLOGIES}


During the preparation of this work, the authors used Claude and Grammarly to improve the readability and language of the manuscript. After using this tool/service, the author(s) reviewed and edited the content as needed and take full responsibility for the content of the published article.

\section*{SUPPLEMENTAL INFORMATION INDEX}




Supplementary material extends the results and methods section from the main article, and it is organized as follows:

\begin{enumerate}
  \item The long-term MARL Electricity Market Model 
  \item The long-term Electricity Market Environment
  \item Multi-Agent Reinforcement Learning Algorithms
  \item Scenarios
  \item Experiments
  \item Independent versus Multi-Agent Proximal Policy Optimization
  \item Additional information and results
\end{enumerate}
\clearpage

\section{The long-term MARL Electricity Market Model}
\label{Appendix - Long-term MARL Electricity Market Model}

This supplementary material highlights the main aspects of the electricity market representation in MARLEY, including the equivalent wholesale market representation and the different long-term regulatory mechanisms that support investments by Generation Companies (GENCOs). 

\subsection{Economic dispatch for equivalent short-term market}
\label{Appendix - Economic dispatch for equivalent short-term market}

The equivalent economic dispatch emulates the short-term operation of the electricity system, providing signals for dispatched resources, prices, and adequacy concerns within the model. To construct the economic dispatch, hourly resource and demand information \cite{antonini_weather-_2024} are aggregated using the TSAM library \cite{hoffmann_typical_2021}, which condenses bi-monthly data into four correlated typical 24-hour days. Once these representative resource and demand series are obtained, one is randomly selected per dispatch call, and the economic dispatch is solved at an hourly resolution over the 24-hour representative period. These dispatches are solved sequentially; each simulated year comprises six economic dispatches, one per bi-monthly period. 

Within the dispatch, four main assumptions are adopted. First, the dispatch minimizes total system costs, as no demand-side flexibility beyond curtailment is considered. Second, generators submit bids based on their marginal production costs, including any applicable carbon taxes. Third, short-term storage is operated to minimize total system costs. Finally, long-term storage \textit{(pumped hydro in the Italian case)} is governed by two rules: during the representative day, it is operated as an 8-hour storage solution, and agents select a target state of charge to be reached by the end of the representative period.

Abiding by these principles, the formulation follows a cost-minimization objective function:
\begin{equation}
\min \sum_{i,t} c_i \cdot q_{i,t} + \sum_{t} c^{voll} \cdot q_{t}^{slack}
\end{equation}
where $c_i$ is the variable cost of generation technology $i$ [\euro{}/MWh], $q_{i,t}$ is the dispatched quantity of technology $i$ at time $t$ [MW], $c^{voll}$ is the Value of Lost Load [\euro{}/MWh], and $q_{t}^{slack}$ is the dispatched quantity of the slack generator at time $t$ [MW], activated only when the system cannot meet demand through available generation and storage resources \textit{(ensuring feasibility inside the Reinforcement Learning loop)}.

The problem is first subject to the supply-demand balance constraint:
\begin{equation}
\sum_{i} q_{i,t} + \sum_{s} (q_{s,t}^{discharge} - q_{s,t}^{charge}) + 
\sum_{n \in N^{l}} (q_{n,l,t}^{discharge} - q_{n,l,t}^{charge}) + q_{t}^{slack} 
= D_t \quad \forall t
\end{equation}
where $D_t$ is the electricity demand at time $t$ [MW], $q_{i,t}$ is the dispatched quantity of generation technology $i$ at time $t$ [MW], $q_{s,t}^{charge}$ and $q_{s,t}^{discharge}$ are the dispatched charge and discharge quantities of short-term storage $s$ at time $t$ [MW], $q_{n,l,t}^{charge}$ and $q_{n,l,t}^{discharge}$ are the dispatched charge and discharge quantities of long-term storage $l$ at time $t$ [MW], corresponding to the operational decision of agent $n \in N^{l}$, where $N^{l}$ is the set of agents holding long-term storage assets, and $q_{t}^{slack}$ is the dispatched quantity of the slack generator at time $t$ [MW].

The dispatched quantity of each generation technology is bounded by its installed capacity, adjusted by a time-varying availability factor to account for the variability of renewable energy sources and the planned or unplanned outages of units \textit{(selected randomly from a pre-defined distribution for each economic dispatch call)}:
\begin{equation}
0 \leq q_{i,t} \leq \bar{Q}_{i,t} \quad \forall i,t
\end{equation}
where $\bar{Q}_{i,t} = \phi_{i,t} \cdot K_i^{max}$ is the available capacity of generation technology $i$ at time $t$ [MW], $K_i^{max}$ is the installed capacity of generation technology $i$ [MW], and $\phi_{i,t} \in [0,1]$ is the availability factor of generation technology $i$ at time $t$.

Short-term ESS are subject to stylized energy-balance dynamics:
\begin{equation}
E_{s,t} = E_{s,t-1} + \eta_{s}^{charge} \cdot q_{s,t}^{charge} - 
\frac{q_{s,t}^{discharge}}{\eta_{s}^{discharge}} \quad \forall s,t
\end{equation}
\begin{equation}
0 \leq E_{s,t} \leq E_s^{max} \quad \forall s,t
\end{equation}
\begin{equation}
Q_s^{min} \leq q_{s,t}^{charge} \leq Q_s^{charge,max} \quad \forall s,t
\end{equation}
\begin{equation}
Q_s^{min} \leq q_{s,t}^{discharge} \leq Q_s^{discharge,max} \quad \forall s,t
\end{equation}
where $E_{s,t}$ is the energy stored in short-term storage $s$ at time $t$ [MWh], $\eta_{s}^{charge}$ and $\eta_{s}^{discharge}$ are the charge and discharge efficiencies of short-term storage $s$, $E_s^{max}$ is the maximum energy capacity of short-term storage $s$ [MWh], $Q_s^{charge,max}$ and $Q_s^{discharge,max}$ are the maximum charge and discharge power capacities of short-term storage $s$ [MW], and $Q_s^{min}$ is the minimum power capacity of short-term storage $s$ [MW]. The previous set of constraints can be repeated for ESS with different energy-to-power ratios.

For long-term ESS, aside from the standard constraints, an additional parameter is included to represent the operational decision taken by incumbent GENCOs for the operation of these assets. These constraints are formulated individually for each agent $n \in N^{l}$ holding long-term storage assets:
\begin{equation}
E_{n,l,t} = E_{n,l,t-1} + \eta_{l}^{charge} \cdot q_{n,l,t}^{charge} - 
\frac{q_{n,l,t}^{discharge}}{\eta_{l}^{discharge}} + I_{l,t} \quad \forall n \in N^{l}, l, t
\end{equation}
\begin{equation}
0 \leq E_{n,l,t} \leq E_{n,l}^{max} \quad \forall n \in N^{l}, l, t
\end{equation}
\begin{equation}
Q_l^{min} \leq q_{n,l,t}^{charge} \leq Q_l^{charge,max} \quad \forall n \in N^{l}, l, t
\end{equation}
\begin{equation}
Q_l^{min} \leq q_{n,l,t}^{discharge} \leq Q_l^{discharge,max} \quad \forall n \in N^{l}, l, t
\end{equation}
\begin{equation}
E_{n,l,T} = \gamma_{n,l} \cdot E_{n,l}^{max} \quad \forall n \in N^{l}, l
\end{equation}
where $E_{n,l,t}$ is the energy stored in long-term storage $l$ at time $t$ by agent $n$ [MWh], $\eta_{l}^{charge}$ and $\eta_{l}^{discharge}$ are the charge and discharge efficiencies of long-term storage $l$, $E_{n,l}^{max}$ is the maximum energy capacity of long-term storage $l$ held by agent $n$ [MWh], $Q_l^{charge,max}$ and $Q_l^{discharge,max}$ are the maximum charge and discharge power capacities of long-term storage $l$ [MW], $Q_l^{min}$ is the minimum power capacity of long-term storage $l$ [MW], $I_{l,t}$ is the water inflow to long-term storage $l$ at time $t$ [MWh], and $\gamma_{n,l}$ is the desired state of charge parameter selected by agent $n$ for long-term storage $l$ at the final period. Importantly, the formulation assumes that all final states in the long-term storage solution for the next representative period are feasible.

Furthermore, the system is subject to a simplified flexibility constraint, aiming to represent network and system constraints in the power system. Future work will focus on improving this representation, either by adding additional constraints or by using aggregated parameters. For the current version:
\begin{equation}
\sum_{i} \alpha_i \cdot q_{i,t} + \sum_{s} \alpha_s \cdot (q_{s,t}^{charge} + 
q_{s,t}^{discharge}) + \sum_{l} \alpha_l \cdot (q_{l,t}^{charge} + 
q_{l,t}^{discharge}) + q_{t}^{slack} \geq \beta \cdot D_t \quad \forall t
\end{equation}
where $\alpha_i$, $\alpha_s$, and $\alpha_l$ are the flexibility coefficients for generation technology $i$, short-term storage $s$, and long-term storage $l$ respectively, and $\beta$ is the flexibility coefficient for demand. The slack generator $q_{t}^{slack}$ is included in the flexibility constraint to ensure the problem's feasibility. 

The flexibility coefficients reflect the degree to which each technology can contribute to system flexibility, with dispatchable and storage technologies generally assigned higher values than variable renewable sources. The parameters used in the model are reported in Table~\ref{tab:flex_coefficients}. We leave for future work an enhancement of the grid module to better represent technical constraints in power systems. 

\begin{table}[h]
\centering
\begin{tabular}{|cc|c|}
\hline
\multicolumn{1}{|c|}{\textbf{Category}}                    & \textbf{Technology} & \textbf{Coefficient} \\ \hline
\multicolumn{1}{|c|}{\multirow{7}{*}{\textbf{Generation}}} & Solar PV            & 0                    \\ \cline{2-3} 
\multicolumn{1}{|c|}{}                                     & Onshore Wind        & 0.1                  \\ \cline{2-3} 
\multicolumn{1}{|c|}{}                                     & Offshore Wind       & 0.12                 \\ \cline{2-3} 
\multicolumn{1}{|c|}{}                                     & Coal                & 0.8                  \\ \cline{2-3} 
\multicolumn{1}{|c|}{}                                     & OCGT                & 0.9                  \\ \cline{2-3} 
\multicolumn{1}{|c|}{}                                     & CCGT                & 0.8                  \\ \cline{2-3} 
\multicolumn{1}{|c|}{}                                     & Others              & 0.1                  \\ \hline
\multicolumn{1}{|c|}{\multirow{2}{*}{\textbf{Storage}}}    & Short-term storage  & 0.8                  \\ \cline{2-3} 
\multicolumn{1}{|c|}{}                                     & Long-term storage   & 0.8                  \\ \hline
\multicolumn{2}{|c|}{\textbf{Demand}}                                            & 0.2                  \\ \hline
\end{tabular}%
\caption{\textbf{Flexibility coefficients for the simplified flexibility constraint.} The values are a stylized representation of flexibility constraints in the power system, based on \cite{najjar_performance_2020,ssengonzi_efficient_2022, pluta_quantifying_2025}}
\label{tab:flex_coefficients}
\end{table}

Finally, the energy price $\lambda_t$ is the shadow price of the supply-demand balance constraint [\euro{}$/MWh$], representing the marginal cost of serving additional electricity demand. The flexibility price $\mu_t$ is the shadow price of the flexibility balance constraint [\euro{}$/MWh$], representing the marginal cost of providing additional flexibility. Based on these definitions, GENCOs are remunerated for the energy dispatched at the energy price, for their flexibility contribution at the flexibility price, and for the specific coefficient of the generation technologies. 

\textbf{Sets and indices:} $i \in I$ (generation technologies), $s \in S$ (short-term storage technologies), $l \in L$ (long-term storage technologies), $t \in T$ (time periods).

\subsection{Contract for Difference Market}
\label{Appendix - Contract for Difference Market}

The model incorporates a Contracts for Difference (CfD) mechanism as a long-term revenue stabilization instrument for renewable energy investment. This section first explains the financial contract, then details the balancing mechanism the policymaker uses to provide the investment signal. 

\subsubsection{Two-way Contract for Differences:}

The CfD operates as a two-way production-based financial contract: when the spot market price exceeds the agreed strike price, the generator pays back the difference to the system. Conversely, when the spot price falls below the strike price, the generator receives a top-up payment. 

The financial settlement of the CfD contract is computed at each time period $t$ of the economic dispatch. From the system's perspective, the net CfD settlement payment is given by:
\begin{equation}
\Phi_{n,j,t}^{CfD} = (\lambda_t - \kappa_{n,j}^{CfD}) \cdot q_{n,j,t}^{CfD} 
\quad \forall n \in N^{CfD}, j \in J^{RES}, t
\end{equation}
where $\lambda_t$ is the spot market price at time $t$ [\euro{}/MWh], $\kappa_{n,j}^{CfD}$ is the CfD strike price awarded to agent $n$ for technology $j$ [\euro{}/MWh], and $q_{n,j,t}^{CfD}$ is the quantity of energy dispatched under the CfD contract by agent $n$ for technology $j$ at time $t$ [MWh]. The share of energy dispatched coming from CfD is defined as the ratio between CfD and total system capacity, so CfD generators have no priority in the dispatch and suffer the direct effects of curtailment. The price applicable to an agent is the weighted average of all assets in a particular technology participating in the market.    

In the case $\Phi_{n,j,t}^{CfD} > 0$, the system receives a net payment from the generator; when $\Phi_{n,j,t}^{CfD} < 0$, the system makes a top-up payment to the generator. 

\subsubsection{RES investment signal:}

The CfD auction is administered by the planner agent \textit{(policymaker)}, which periodically assesses the gap between projected and target renewable energy penetration over a planning horizon. Specifically, the regulator projects the future RES energy contribution \textit{(accounting for existing capacity, assets under construction, and capacity aging)} and compares it against a target penetration level, itself determined by the planner's action in the previous period. An auction is triggered only when the projected RES penetration falls short of the target, ensuring that the mechanism intervenes only when a genuine supply gap exists. A single pay-as-bid auction is held jointly for the three RES technologies in the model \textit{(solar PV, onshore wind, and offshore wind)} in which generators submit price and quantity bids, and winning projects are awarded a CfD contract at their bid price. Accepted projects are subsequently added to the investment pipeline and built following their respective construction times.

At each decision step, the planner evaluates the current RES balance by comparing the projected average yearly RES energy contribution against the projected average yearly demand over a planning horizon $H$:
\begin{equation}
\hat{\rho}_{t+H} = \frac{\hat{E}_{t+H}^{RES}}{\hat{D}_{t+H}}
\end{equation}
where $\hat{E}_{t+H}^{RES}$ is the projected average yearly RES energy contribution at the planning horizon [MWh], accounting for existing installed capacity, all projects currently under construction, and expected capacity aging, and $\hat{D}_{t+H}$ is the projected average yearly electricity demand at the planning horizon [MWh]. The planner then decides whether the current projected penetration $\hat{\rho}_{t+H}$ should be maintained or increased, by selecting an incremental target:
\begin{equation}
\rho^{target}_{t} = \hat{\rho}_{t+H} + \delta^{CfD}_{t}
\end{equation}
where $\delta^{CfD}_{t} \geq 0$ is the incremental penetration target selected by the planner at decision step $t$, drawn from a discrete action space. When $\delta^{CfD}_{t} = 0$, no auction is triggered and the current deployment trajectory is deemed sufficient. When $\delta^{CfD}_{t} > 0$, the resulting RES balance gap:
\begin{equation}
\Delta_{t}^{RES} = \left(\hat{\rho}_{t+H} + \delta^{CfD}_{t}\right) \cdot 
\hat{D}_{t+H} - \hat{E}_{t+H}^{RES} \quad \forall t
\end{equation}
is strictly positive, triggering a CfD auction for a contracted volume equal to $\Delta_{t}^{RES}$. 

GENCO agents submit price and quantity bids for solar PV, onshore wind, and offshore wind in a single pay-as-bid auction, with winning projects awarded a CfD strike price equal to their submitted bid. Price bids are drawn from a discrete action space ranging from $0$ to the auction price cap $\bar{\kappa}^{CfD}$ [\euro{}/MWh], following a multi-discrete distribution over a fixed number of price steps. Quantity bids are expressed as the average yearly energy production of the proposed plant [MWh], computed as the product of the installed capacity and the average yearly availability factor of the corresponding technology. The investment capacity itself is selected from a set of discrete investment steps $\mathcal{K} = \{K^1, K^2, \ldots, K^M\}$ [MW], where the steps are designed to provide finer granularity at lower capacity levels.

All accepted projects, including the marginal bid in the auction, are subsequently added to the investment pipeline and built following the construction time of the corresponding technology.

\subsection{Short-term flexibility market}
\label{Appendix - Short-term flexibility market}

The model incorporates a mechanism to support the integration of short-term storage. As before, this section explains the financial contract first, and later details the balancing mechanism used by the policymaker to provide the investment signal. 

\subsubsection{Financial support for flexibility:}

Generators awarded a flexibility contract receive a fixed periodic capacity payment for their installed energy capacity. This payment is independent of the actual dispatch outcome and is settled at each representative period. The capacity payment received by agent $n$ for storage technology $s$ is given by:
\begin{equation}
\Phi_{n,s}^{FL} = \kappa_{n,s}^{FL} \cdot E_{n,s}^{FL} \cdot \omega
\quad \forall n \in N^{FL}, s
\end{equation}
where $\kappa_{n,s}^{FL}$ is the awarded capacity payment for agent $n$ and storage technology $s$ [\euro{}/MWh of installed energy capacity], $E_{n,s}^{FL}$ is the contracted energy capacity [MWh], and $\omega$ is the number of hours in the representative period.

In exchange for this guaranteed revenue stream, generators are awarded flexibility contracts in which they forgo the flexibility remuneration arising from the economic dispatch. Specifically, while all storage assets contribute to the flexibility constraint described in the dispatch formulation, resources holding a flexibility contract do not receive the flexibility price $\mu_t$ [\euro{}/MWh] for their flexibility contribution. The foregone flexibility revenue for agent $n$ and storage technology $s$ at time $t$ is:
\begin{equation}
\Psi_{n,s,t}^{FL} = \mu_t \cdot \alpha_s \cdot 
(q_{n,s,t}^{charge} + q_{n,s,t}^{discharge}) \cdot 
\frac{E_{n,s}^{FL}}{E_{n,s}^{total}}
\quad \forall n \in N^{FL}, s, t
\end{equation}
where $\mu_t$ is the flexibility price at time $t$ [\euro{}/MWh], $\alpha_s$ is the flexibility coefficient of storage technology $s$, $q_{n,s,t}^{charge}$ and $q_{n,s,t}^{discharge}$ are the dispatched charge and discharge quantities of storage $s$ at time $t$ [MW], and $E_{n,s}^{FL}/E_{n,s}^{total}$ is the share of agent's storage capacity under the flexibility contract relative to its total installed storage capacity. 

\subsubsection{Flexibility investment signal:}

Similar to the CfD mechanism, the flexibility auction is administered by the planner agent, which periodically assesses the adequacy of short-term storage capacity relative to system demand over 
a planning horizon. Specifically, the policymaker evaluates the current ratio of installed short-term storage energy capacity against projected average yearly demand, and intervenes when this ratio falls short of a target level. A single pay-as-bid auction is held for short-term storage technologies, in which generators submit price and quantity bids, and winning projects are awarded a capacity payment contract at their bid price. Accepted projects are subsequently added to the investment pipeline and built following their respective construction times.

At each decision step, the planner evaluates the current storage adequacy ratio by comparing the projected short-term storage energy capacity against the projected average yearly demand over a planning horizon $H$:
\begin{equation}
\hat{\sigma}_{t+H} = \frac{\hat{E}_{t+H}^{ST}}{\hat{D}_{t+H}}
\end{equation}
where $\hat{E}_{t+H}^{ST}$ is the projected short-term storage energy capacity at the planning horizon [MWh], accounting for existing installed capacity, all projects currently under construction, and expected capacity aging, and $\hat{D}_{t+H}$ is the projected average yearly electricity demand at the planning horizon [MWh].

The planner then decides whether the current projected ratio $\hat{\sigma}_{t+H}$ should be maintained or increased, by selecting an incremental target:
\begin{equation}
\sigma^{target}_{t} = \hat{\sigma}_{t+H} + \delta^{FL}_{t}
\end{equation}
where $\delta^{FL}_{t} \geq 0$ is the incremental storage adequacy target selected by the planner at decision step $t$, drawn from a discrete action space. When $\delta^{FL}_{t} = 0$, no auction is triggered and the current storage deployment trajectory is deemed sufficient. When $\delta^{FL}_{t} > 0$, the resulting storage balance gap:
\begin{equation}
\Delta_{t}^{FL} = \left(\hat{\sigma}_{t+H} + \delta^{FL}_{t}\right) \cdot 
\hat{D}_{t+H} - \hat{E}_{t+H}^{ST} \quad \forall t
\end{equation}
is strictly positive, triggering a flexibility auction for a contracted volume equal to $\Delta_{t}^{FL}$.

GENCO agents submit price and quantity bids for short-term storage technologies in a single pay-as-bid auction, with winning projects awarded a capacity payment at their submitted bid price [\euro{}/MWh of installed energy capacity]. Price bids are drawn from a discrete action space ranging from $0$ to the auction price cap $\bar{\kappa}^{FL}$ [\euro{}/MWh], following a multi-discrete distribution over a fixed number of price steps. Quantity bids are expressed in terms of the energy capacity of the proposed storage plant [MWh]. The investment capacity is selected from the same set of discrete investment steps $\mathcal{K}$ as in the CfD auction, providing finer granularity at lower capacity levels. 

All accepted projects, including the marginal bid in the auction, are subsequently added to the investment pipeline and built following the construction time of the corresponding technology.

\subsection{Capacity Remuneration Mechanism}
\label{Appendix - Capacity Remuneration Mechanism}

For the capacity remuneration mechanism, a reliability option is implemented. This is explained first, while the balancing mechanism, based on the Expected Unserved Energy, is detailed later. 

\subsubsection{Reliability option:}

In economic dispatch, the capacity market serves as a reliability option. Generators awarded a capacity contract receive a continuous capacity payment for their installed capacity, but are required to return to the system any spot market revenues exceeding a pre-defined strike price $\bar{\lambda}^{CM}$ [\euro{}/MWh]. The net capacity market settlement for agent $n$ and technology $j$ at time $t$ is:
\begin{equation}
\Phi_{n,j,t}^{CM} = \kappa_{n,j}^{CM} \cdot K_{n,j}^{CM} \cdot \omega - 
\max(\lambda_t - \bar{\lambda}^{CM}, 0) \cdot K_{n,j}^{CM}
\quad \forall n \in N^{CM}, j, t
\end{equation}
where $\kappa_{n,j}^{CM}$ is the awarded capacity payment [\euro{}/MW], $K_{n,j}^{CM}$ is the contracted capacity of agent $n$ for technology $j$ [MW], $\omega$ is the number of hours in the representative period, $\lambda_t$ is the spot market price at time $t$ [\euro{}/MWh], and $\bar{\lambda}^{CM}$ is the strike price of the reliability option [\euro{}/MWh]. When the spot price exceeds the strike, the generator returns the excess revenues to the system. 

Currently, the strike price is fixed at a value much lower than the system's Value of Lost load \textit{(at 300 [\euro{}/MWh], compared with 4000 [\euro{}/MWh])}, ensuring demands receive benefits during scarcity events and guaranteeing generators cover their variable production costs. 

\subsubsection{Capacity Remuneration Mechanism investment signal:}
The capacity market auction is administered by the planner agent, which periodically assesses whether the current generation and storage portfolio is sufficient to meet a target reliability standard, expressed as a maximum allowable Expected Unserved Energy (EUE). The adequacy assessment is carried out across the full set of representative periods used to describe the operating year, \textit{including} the maximum-demand representative period. Each representative condition $s$ spans the six bi-monthly seasons at hourly resolution and is assigned a probability $p_s$ reflecting its frequency in the sampling distribution, with the extreme maximum-demand condition carrying a small probability \textit{(here $1.7\%$)}. The system EUE is then evaluated as the probability-weighted expectation of the per-condition EUE.

For a given representative condition $s$, the EUE is computed by solving a reliability Linea Program that minimizes weighted unserved energy across all representative hours, subject to generation limits, storage dynamics (short- and long-term), and supply--demand balance:
\begin{equation}
\text{EUE}_s = \sum_{\tau} \omega_{\tau} \cdot u_{s,\tau}
\end{equation}
where $u_{s,\tau}$ is the unserved energy at hour $\tau$ of condition $s$ [MWh] and $\omega_{\tau}$ is the weight of hour $\tau$, which scales a representative day to the duration of its bi-monthly season \textit{(uniform across hours and seasons)}. The system-level EUE used by the planner at decision step $t$ is the expectation over representative conditions:
\begin{equation}
\text{EUE}_t = \mathbb{E}_s\!\left[\text{EUE}_s\right] = \sum_{s} p_s \, \text{EUE}_s .
\end{equation}
The reliability target is defined as a fraction of the expected total energy demand:
\begin{equation}
\text{EUE}^{target} = \epsilon \cdot \mathbb{E}_s\!\left[D_s^{tot}\right],
\qquad D_s^{tot} = \sum_{\tau} \omega_{\tau}\, d_{s,\tau},
\end{equation}
where $\epsilon$ is the maximum allowable EUE fraction and $D_s^{tot}$ is the total (weighted) energy demand of condition $s$ [MWh].

An auction is triggered when the expected system EUE exceeds the target. The volume to be contracted is obtained by converting the EUE gap [MWh] into firm capacity [MW] using the marginal EUE reduction delivered by perfect firm capacity. Defining the expected marginal reduction per MW of perfect capacity and the contract volume as: 

\begin{equation}
\rho = \frac{1}{\Delta K}\,\mathbb{E}_s\!\left[\text{EUE}_s(\mathbf{K}) - \text{EUE}_s(\mathbf{K} + \Delta K \cdot \mathbf{e}^{perfect})\right],
\end{equation}

\begin{equation}
\Delta_t^{CM} = \max\!\left(0,\ \frac{\text{EUE}_t - \text{EUE}^{target}}{\rho}\right).
\end{equation}

As such, $\Delta_t^{CM}$ [MW] represents the additional firm capacity equivalent needed to restore the system to the reliability standard, and an auction is called only when $\Delta_t^{CM} > 0$.

To participate in the auction, the contribution of different resources to system adequacy \textit{(deratings)} is required. For this, the Capacity Remuneration Mechanism computes the marginal ELCC of technology $j$, which measures the fraction of perfect firm capacity that an additional unit of that technology can displace while maintaining the same EUE level. Consistent with the multi-condition assessment, the ELCC is formed as the ratio of probability-weighted EUE reductions, aggregating numerator and denominator across conditions before dividing:

\begin{equation}
\psi_j = \frac{\mathbb{E}_s\!\left[\text{EUE}_s(\mathbf{K}) - \text{EUE}_s(\mathbf{K} + \Delta K \cdot \mathbf{e}_j)\right]}
{\mathbb{E}_s\!\left[\text{EUE}_s(\mathbf{K}) - \text{EUE}_s(\mathbf{K} + \Delta K \cdot \mathbf{e}^{perfect})\right]} \quad \forall j ,
\end{equation}

where $\mathbf{K}$ is the vector of installed capacities, $\mathbf{e}_j$ is a unit vector in the direction of technology $j$, and $\mathbf{e}^{perfect}$ is the unit vector corresponding to the ideal generator. Aggregating the expectations separately ensures that high-stress conditions \textit{(typically the maximum-demand day)} dominate both numerator and denominator, so that a technology's reliability contribution is not diluted by mild conditions in which no unserved energy occurs.

When the adequacy check finds the system already compliant \textit{(i.e. $\Delta_t^{CM} = 0$ and no auction is triggered)}, the model instead reports each technology's average availability factor as its capacity credit. These values are provided only for completeness, since no auction takes place and the credits play no allocative role in that step.

Once an auction is called, the GENCO agents submit price and quantity bids in a pay-as-bid auction, where quantity bids are weighted by the ELCC of the corresponding technology, so that the auctioned volume is expressed in firm capacity equivalents. Accepted projects are subsequently added to the investment pipeline and built following their respective construction times.

\subsubsection{Treatment of existing assets:}
The adequacy assessment and the auction clearing described above account for the full operating portfolio: all existing assets not scheduled for decommissioning are included in evaluating the system EUE and in computing the technology deratings $\psi_j$. In this sense, the firm capacity already provided by the incumbent fleet is fully credited toward the reliability standard, and the contract volume $\Delta_t^{CM}$ reflects only the residual adequacy gap that the incumbent portfolio cannot cover.

Existing assets, however, are remunerated under a distinct rule rather than bidding actively in the auction. In any step where an adequacy gap is identified \textit{(i.e. $\Delta_t^{CM} > 0$)}, each non-decommissioned existing asset receives a capacity premium equal to half of the average premium awarded to new entrants across all technologies:

\begin{equation}
\kappa_{n,j}^{CM,\,exist} =
\begin{cases}
\tfrac{1}{2}\,\bar{\kappa}^{CM,\,new} & \text{if } \Delta_t^{CM} > 0, \\[4pt]
0 & \text{otherwise,}
\end{cases}
\quad \forall n,j,t,
\end{equation}

where $\bar{\kappa}^{CM,\,new}$ is the average premium cleared in the auction for new capacity across all accepted technologies. When no adequacy gap is identified and no auction is triggered, existing assets receive no capacity premium. As with new entrants, existing assets remain subject to the reliability-option settlement of Equation~(1), returning to the system any spot revenues above the strike price $\bar{\lambda}^{CM}$.

This treatment reflects the modelling assumption that incumbent capacity contributes to adequacy and is partially remunerated for doing so during scarcity, while the marginal investment signal is reserved for new entrants. In future work, we plan to relax this assumption by allowing existing assets to bid actively in the capacity auction.

\clearpage

\section{The Long-term Electricity Market Environment}
\label{Appendix - Long-term Electricity Market Environment}

The Long-term Electricity Market is implemented following the Gymnasium standard \cite{towers_gymnasium_2023}, and more specifically the multi-agent environment interface developed by the RLlib team \cite{liang_rllib_2018}. This supplementary material describes that implementation.

\subsection{Environment Structure}

In the Gymnasium standard, reinforcement learning environments advance in discrete steps that simulate transitions in the underlying Markov Decision Process. At each step, agents observe the system state, take actions, and receive the corresponding reward. In the market model, these environment steps map directly onto Equivalent Short-Term Market sessions and their associated Representative Periods.

During each step, the GENCOs' portfolios participate in the Equivalent Short-Term Market, producing the outcomes that form the basis for the environment's observations and rewards. Aside from operating mid-term storage, however, agents do not make decisions that affect the short-term market: generation assets bid their availability and marginal costs to the day-ahead market (strategic bidding is not enabled), and the intra-day charge and discharge cycles of the storage systems are scheduled for efficient system operation. Investment decisions, by contrast, are enabled yearly, allowing agents to expand their portfolios. These investments occur every two environment steps and in sequence across the year, as shown in the lower panel of Figure \ref{fig: General Overview}, so that yearly investment in any market is possible whenever system conditions require it.

\subsection{GENCO}

\subsubsection{Reward}

GENCOs are assumed to maximize the net present value of the cash flows obtained from the electricity market, aggregating revenues and costs across all assets and markets in their portfolio over the simulation. To remain consistent with real company operations, the reward is delivered to agents step by step, as in Equation (\ref{eq:GENCO - total_reward}). It is split into two parts: the profits $(P_{\mathrm{M}}, P_{\mathrm{CM}}, P_{\mathrm{CfD}})$ earned across the system's markets, and the investment costs $(IC_{\mathrm{M}}, IC_{\mathrm{CM}}, IC_{\mathrm{CfD}})$ of new portfolio additions, disaggregated by the same markets. To keep the financial modeling transparent, cash-flow discounting is performed within the environment, with the discount rate treated as an exogenous parameter.

\begin{equation}
R_{t}^{g} = \Big( \underbrace{P_{\mathrm{M}} + P_{\mathrm{CM}} + P_{\mathrm{CfD}}}_{\text{market profits}}
\;-\; \underbrace{\big(IC_{\mathrm{M}} + IC_{\mathrm{CM}} + IC_{\mathrm{CfD}}\big)}_{\text{investment costs}} \Big)\,(1+r_{g})^{-t}
\label{eq:GENCO - total_reward}
\end{equation}

\subsubsection{Action Space}

GENCO actions comprise: (i) discrete merchant investments; (ii) quantity and price bids in the capacity market; (iii) quantity and price bids in the Contract-for-Difference market; and (iv) operating decisions for the mid-term storage assets.

\subsubsection{Observation Space}

The set of observations available to GENCOs follows three principles. First, the selected variables should reflect real market conditions, in which agents access publicly shared system information while firm-specific details remain private and inaccessible to competitors. Second, no internal price-forecasting tools are included, preserving the agents' full autonomy in decision-making. Third, any information available in the market is also provided to the agents.

Accordingly, the GENCO observation set includes indicative time series for demand and variable-resource availability, the energy mix implied by the assets in operation, individual and system-wide reservoir levels, market information \textit{(such as prices and balances from the Capacity and CfD markets)}, the aggregated reward per technology in those markets, policy-relevant information, and the time index of the environment step. Because it relies on aggregated system information, the observation framework does not depend on the number of agents, which is constant within each simulation.

\subsection{Policymaker}

\subsubsection{Reward}

The policymaker's reward is a function of three components, as shown in Equation (\ref{eq:Policymaker - total_reward}). First, total electricity system costs $EC_{t}$ cover the energy dispatched in the short-term sessions plus any settlements from the financial mechanisms tied to the investment channels. Second, the emissions cost $C_{\mathrm{CO_2},t}$ values the system emissions $E_{\mathrm{CO_2},t}$ at a time-dependent Social Cost of Carbon $\mathit{SCC}_{t}$, so that $C_{\mathrm{CO_2},t} = \mathit{SCC}_{t}\,E_{\mathrm{CO_2},t}$. We set the trajectory $\mathit{SCC}_{t}$ to coincide with the carbon price path applied in each scenario, so that the policymaker internalizes emissions at the same per-tonne value that GENCOs face through carbon pricing. This avoids a mismatch between the cost imposed on emitters and the cost perceived by the policymaker. Third, the term $Tax_{t}$ collects the revenues from carbon pricing and corporate taxes. As for GENCOs, discounting is performed within the environment, using a discount rate $r_{p}$ specific to the policymaker.

\begin{equation}
R_{t}^{p} = \big( -\,EC_{t} - \mathit{SCC}_{t}\,E_{\mathrm{CO_2},t} + Tax_{t} \big)\,(1+r_{p})^{-t}
\label{eq:Policymaker - total_reward}
\end{equation}
\subsubsection{Action Space}

The policymaker takes key decisions in the markets where GENCOs interact: in the CfD and flexibility markets, it sets the RES and storage penetration targets. As with the GENCOs, the policymaker's actions are modeled through discrete variables, and their activation is governed by Action Masking.

\subsubsection{Observation Space}

The policymaker observation set follows the same logic as GENCO's, receiving indicative time series for demand and variable-resource availability, the energy mix implied by the assets in operation, individual and system-wide reservoir levels, and market information \textit{(such as prices and balances from the Capacity and CfD markets)}. Instead of economic profit, however, the reward observations disaggregate the terms shown in Equation (\ref{eq:Policymaker - total_reward}).

\subsection{Initial and Absorbing States}
\label{Appendix - Final state}

Two boundary effects require special treatment. At the start, the environment cannot represent a pre-existing construction pipeline; this is approximated by holding demand and the generation mix stable for the first few years, allowing agents to build their own pipeline of projects under construction.

At the end, initial tests showed that agents avoid investing in assets that would not be compensated for the remaining operational lifetime beyond the horizon. To correct this, the absorbing state introduces a terminal payment $P_{\mathrm{absorbing}}$ (Equation \ref{eq:annuity absorbing state}): the present value of an average income $I_{\mathrm{mean}}$ over each asset's remaining lifetime $t_{\mathrm{rem}}$. The average income is approximated by the system-wide mean profit and computed separately for each technology and market to reflect portfolio heterogeneity.

\begin{equation}
P_{\mathrm{absorbing}} = I_{\mathrm{mean}}\,\frac{1 - (1+r)^{-t_{\mathrm{rem}}}}{r}
\label{eq:annuity absorbing state}
\end{equation}

Two further safeguards smooth the simulation's end: agents cannot start investments that would not become operational before termination, and demand and the generation mix are again held stable to avoid artificial scarcity or oversupply. These boundary adjustments improve stability at the cost of additional simulation and training steps. Overall, for the current setup that analyzes the 2025-2040 horizon, we simulate 30 years of operation. 

\clearpage

\section{Multi-Agent Reinforcement Learning Algorithms}

This supplementary material extends the discussion of the two multi-agent reinforcement learning algorithms explored in this work: Independent Learning and Proximal Policy Optimization (IPPO), and Multi-Agent Proximal Policy Optimization (MAPPO) 

Starting from the single-agent baseline, Proximal Policy Algorithm (PPO) is an Actor-Critic On-Policy algorithm introduced by \citet{schulman_proximal_2017}. Similar to other Actor-Critic On-Policy algorithms, agents in PPO aim to maximize a cumulative reward function over a Markov Decision Process (MDP). Assuming partial observability, trajectories in an MDP are a succession of observations, actions, and rewards,  \( \tau = (o_0, a_0, r_0, \dots o_t, a_t, r_t, \dots, o_T, a_T, r_T) \), where the subscript \(t\) represents time steps, and \([0, T]\) denote the initial and terminal steps in the system \cite{sutton_reinforcement_2020}. 

To achieve this objective, PPO trains a stochastic policy via two neural networks. On the one hand, the Critic, denoted by \( V_{w}(o_{t}) \) and trainable parameters \(w\), is in charge of providing a baseline for the Reward estimation. On the other hand, the Actor, defined as \( \pi_{\theta_(a_{t} | o_{t})} \) and parameters \( \theta\), selects reward-maximizing actions based on the current observed state. Training follows a three-part Loss: a conservative update to the Actor's policy when selecting actions via the PPO Clipping, an Entropy term that incentivizes exploration, and a squared loss related to the Reward regression problem in the Critic \cite{schulman_proximal_2017,bick_towards_2021}. 

Following the extension of PPO towards multi-agent cooperative and competitive tasks introduced in \citet{yu_surprising_2022}, IPPO replicates the training process to every agent in the Partially Observable Stochastic Game (POSG). In particular, the POSG considers a set of $N$ agents, $I = \{1,\dots, i, \dots, N\}$, each with its set of actions \(a_{i,t}\) and receiving a subset of observations  \(o_{i,t}\) from the joint environment. For each agent, IPPO trains independent Actor and Critic networks, \( \pi_{i,\theta_{i}}(a_{i,t} | o_{i,t}) \) and \( V_{i,w_i}(o_{i,t}) \), for agent \(i \) \cite{albrecht_multi-agent_2024}. Following \citet{gonzalez-ruiz_assessing_2026}, training these networks uses expressions [\ref{eq: objective IPPO}-\ref{eq: Vf target IPPO}], where:

\begin{equation}
\label{eq: objective IPPO}
L_{i}(\theta_{i}, w_{i}) = L_{i}^{\text{CLIP}}(\theta_{i}) + h_{i} L_{i}^{\text{entropy}}(\theta_{i}) - v_{i} L_{i}^{\text{VF}}(w_{i})
\end{equation}

\begin{equation}
\label{eq: clip IPPO}
L_{i}^{\text{CLIP}}(\theta_{i}) = \hat{\mathbb{E}_{i,t}}\left[ \min \left( p_{i,t}(\theta_{i}) \hat{A_{i,t}}, \text{clip} \left(p_{i,t}(\theta_{i}), 1 - \epsilon_{i}, 1 + \epsilon_{i}\right) \hat{A_{i,t}} \right) \right]
\end{equation}

\begin{equation}
\label{eq: Advantage IPPO}
\hat{A_{i,t}} =V_{i,t}^{target} - V_{i,w_{i,old}}(o_{i,t})
\end{equation}

\begin{equation}
\label{eq: Vf IPPO}
 L_{i}^{\text{VF}}(w_{i}) = \hat{\mathbb{E}_t} \left[ \left( V_{i,w_i}(o_{i,t}) - V_{i,t}^{target} \right)^2 \right]
\end{equation}

\begin{equation}
\label{eq: Vf target IPPO}
V_{i,t}^{target} = r_{i,t} + \gamma_i r_{i,t+1} + \gamma_i^2 r_{i,t+2} + \dots + \gamma_i^{n-1} r_{i,t+n-1} +\gamma_i^{n} V_{i,w_{i,old}}(o_{i,t+n})
\end{equation}

\begin{itemize}
\item The terms \( L_{i}^{\text{CLIP}}(\theta_{i})\) and \(L_{i}^{\text{entropy}}(\theta_{i}) \) guide the updates in the Actor network \( \pi_{i,\theta_{i}}(a_{i,t} | o_{i,t}) \), while \(L_{i}^{\text{VF}}(w_{i})\) drive updates in the Critic network \( V_{i,w_i}(o_{i,t}) \); 
\item \(\theta_{i}\) and \(w_i\) are the parameters of the Actor and Critic networks. Furthemore, parameters from the previous algorithm iteration are referred to as \(\theta_{i,old}\) and \(w_{i,old}\);
\item The main objective function of PPO, \( L_{i}^{\text{CLIP}}(\theta_{i})\), uses an Advantage estimation to shift the Actor Policy toward actions that maximize the expected reward while controlling for the maximum size in the updates using the combination of the \( \text{clip} \) and \(\min \) functions;
\item \( \epsilon_{i} \) is a hyperparameter that, in conjunction with the clipping function, limits the ratio between new and old policies to remain in the range \([1 - \epsilon_{i}, 1 + \epsilon_{i}]\);
\item \( L_{i}^{\text{entropy}}(\theta_{i}) \) is an entropy term that procures the exploration of new strategies by inducing randomness in action selection in the Actor network;
\item \( p_{i,t}(\theta_{i}) \) is defined as the variation in actor policies between the algorithm updates, measured by the fraction \( \frac{\pi_{\theta_{i}}(a_{i,t} | o_{i,t})}{\pi_{\theta_{i,old}}(a_{i,t} | o_{i,t})} \); 
\item \(A_{i,t}\) corresponds to the Advantage function, estimated via the difference between an estimate of the Action-state function, \(V_{i,t}^{target}\), denoted as target state value, and the previous version of the Value-state function \(V_{w_{i,old}}(o_{i,t})\);
\item \( L_{i}^{\text{VF}}(w_{i})\) drives the updates in the Critic network via minimizing the squared error between the predicted State-value function \(V_{w_{i}}(o_{i,t}) \), and the estimated target state value \(V_{i,t}^{target} \);
\item \(V_{i,t}^{target}\) is used as a proxy of the Action-state value function, and is calculated using the accumulated rewards in a given trajectory with length \((t+n)\); and
\item \( h_{i} \) and \( v_{i} \) are coefficients that balance exploration, via the entropy term, and the weight of the value function loss with respect to the total loss. 
\end{itemize}

The training cycle starts with initializing the Actor and Critic networks. Using the current version of the Actor Policy, trajectories are collected from the environment until a given batch is filled. Using these trajectories, the networks are updated by applying the previous loss functions and multiple epochs of Stochastic Gradient Descent. This process is repeated until convergence, or until the training budget is reached. Once agents are trained, the actions are sampled exclusively from the Critic, using the individual agent observations to feed the neural network. 

In contrast to IPPO, where each agent learns independently from its local observations, MAPPO adopts the centralized-training-decentralized-execution (CTDE) paradigm introduced by \citet{yu_surprising_2022}, inspired by similar architectures used in other multi-agent algorithms \cite{lowe_multi-agent_2017}. Over the same POSG with $N$ agents, $I = \{1,\dots, i, \dots, N\}$, each agent retains an independent Actor network \( \pi_{i,\theta_{i}}(a_{i,t} | o_{i,t}) \) conditioned on its local observation \(o_{i,t}\), preserving decentralized execution. The value estimation is performed by a single Critic whose parameters \(w\) are shared across all agents, while a dedicated value head with parameters \(\psi_i\) produces the estimate for each agent \(i\). The Critic is evaluated on a centralized state \(s_{t}\) that aggregates all agents' observations. To accommodate the heterogeneous reward scales across agents, the per-agent value targets are normalized using running per-agent statistics, as proposed by \citet{yu_surprising_2022}. Training these networks uses expressions [\ref{eq: objective MAPPO}-\ref{eq: Vf target MAPPO}], where:

\begin{equation}
\label{eq: objective MAPPO}
L(\theta_{i}, w, \psi_{i}) = L_{i}^{\text{CLIP}}(\theta_{i}) + h_{i} L_{i}^{\text{entropy}}(\theta_{i}) - v_{i} L_{i}^{\text{VF}}(w, \psi_{i})
\end{equation}

\begin{equation}
\label{eq: clip MAPPO}
L_{i}^{\text{CLIP}}(\theta_{i}) = \hat{\mathbb{E}_{i,t}}\left[ \min \left( p_{i,t}(\theta_{i}) \hat{A_{i,t}}, \text{clip} \left(p_{i,t}(\theta_{i}), 1 - \epsilon_{i}, 1 + \epsilon_{i}\right) \hat{A_{i,t}} \right) \right]
\end{equation}

\begin{equation}
\label{eq: Advantage MAPPO}
\hat{A_{i,t}} = V_{i,t}^{target} - V_{i,w_{old},\psi_{i,old}}(s_{t})
\end{equation}

\begin{equation}
\label{eq: denorm MAPPO}
V_{i,w,\psi_{i}}(s_{t}) = \mu_{i} + \sigma_{i}\, \tilde{V}_{i,w,\psi_{i}}(s_{t})
\end{equation}

\begin{equation}
\label{eq: Vf MAPPO}
L_{i}^{\text{VF}}(w, \psi_{i}) = \hat{\mathbb{E}}_t \left[ \left( \tilde{V}_{i,w,\psi_{i}}(s_{t}) - \frac{V_{i,t}^{target} - \mu_{i}}{\sigma_{i}} \right)^2 \right]
\end{equation}

\begin{equation}
\label{eq: Vf target MAPPO}
V_{i,t}^{target} = r_{i,t} + \gamma_i r_{i,t+1} + \gamma_i^2 r_{i,t+2} + \dots + \gamma_i^{n-1} r_{i,t+n-1} + \gamma_i^{n} V_{i,w_{old},\psi_{i,old}}(s_{t+n})
\end{equation}

\begin{itemize}
\item The terms \( L_{i}^{\text{CLIP}}(\theta_{i})\) and \(L_{i}^{\text{entropy}}(\theta_{i}) \) guide the updates in the Actor network \( \pi_{i,\theta_{i}}(a_{i,t} | o_{i,t}) \), while \(L_{i}^{\text{VF}}(w, \psi_{i})\) drives updates in the shared Critic body and the value head of agent \(i\);
\item \(\theta_{i}\) are the parameters of the Actor network of agent \(i\); \(w\) denotes the parameters of the Critic body shared across all agents, and \(\psi_{i}\) the parameters of the value head dedicated to agent \(i\). Parameters from the previous algorithm iteration are referred to as \(\theta_{i,old}\), \(w_{old}\), and \(\psi_{i,old}\);
\item \( s_{t} \) is the centralized state, common to all agents;
\item \( \tilde{V}_{i,w,\psi_{i}}(s_{t}) \) is the raw output of the value head of agent \(i\), expressed in the normalized value space, whereas \( V_{i,w,\psi_{i}}(s_{t}) \) is the corresponding denormalized estimate recovered through expression [\ref{eq: denorm MAPPO}]; \( \mu_{i} \) and \( \sigma_{i} \) are the running mean and standard deviation of the value targets of agent \(i\), as implemented in \citet{yu_surprising_2022};
\item \( \epsilon_{i} \) is a hyperparameter that, in conjunction with the clipping function, limits the ratio between new and old policies to remain in the range \([1 - \epsilon_{i}, 1 + \epsilon_{i}]\);
\item \( L_{i}^{\text{entropy}}(\theta_{i}) \) is an entropy term that procures the exploration of new strategies by inducing randomness in action selection in the Actor network;
\item \( p_{i,t}(\theta_{i}) \) is defined as the variation in actor policies between the algorithm updates, measured by the fraction \( \frac{\pi_{\theta_{i}}(a_{i,t} | s_{t})}{\pi_{\theta_{i,old}}(a_{i,t} | s_{t})} \);
\item \(A_{i,t}\) corresponds to the Advantage function, estimated via the difference between the target state value \(V_{i,t}^{target}\) and the previous version of the denormalized value of agent \(i\), \(V_{i,w_{old},\psi_{i,old}}(s_{t})\);
\item \( L_{i}^{\text{VF}}(w, \psi_{i})\) drives the updates in the shared Critic via minimizing, in the normalized value space, the squared error between the raw value-head output \(\tilde{V}_{i,w,\psi_{i}}(s_{t}) \) and the normalized target \( (V_{i,t}^{target} - \mu_{i})/\sigma_{i} \); and
\item \( h_{i} \) and \( v_{i} \) are coefficients that balance exploration, via the entropy term, and the weight of the value function loss with respect to the total loss.
\end{itemize}

The training cycle is carried out similarly to the IPPO case, with the difference that the Critic updates are computed using centralized information collected from all agents' local observation sets. Likewise, agents' actions are sampled for each Actor network using only the local observation set available to each agent. 

Given the previous algorithms, it is worth discussing the advantages and disadvantages of each, referring the reader to \citet{albrecht_multi-agent_2024} for an in-depth discussion of the different multi-agent archetypes. Starting with IPPO, it is the simpler implementation, requiring only straightforward changes to the PPO algorithm and fewer tunable parameters to control for the multi-agent setting. In competitive environments, IPPO avoids information sharing between agents, which is more consistent with markets where no information is actually shared between competitors. It also allows individual agents to develop their own strategies while remaining less dependent on shared parameters. Finally, for practical purposes, when agent observations are relatively large, IPPO scales more favorably: because each agent owns a separate critic that conditions only on its own observation, adding agents leaves the per-agent network size unchanged and increases total memory only linearly, with no single network growing with the agent count, given that training at the GPU level is done in parallel on a per-agent basis. However, IPPO offers no guarantees against credit-assignment issues, which increases the difficulty for each agent's learning. Particularly, the algorithm cannot distinguish between reward variations caused by changes in other agents' strategies and those caused by its own actions. As a result, each Actor faces a non-stationary environment that violates the main assumptions underlying Reinforcement Learning, thereby preventing convergence to a local optimum.

In contrast, the CTDE paradigm enables MAPPO to reduce the non-stationarity faced by each Actor, since value estimation is grounded on the global state, while still allowing each head to specialize in the value function of its agent \cite{lowe_multi-agent_2017}. This implementation also reduces the number of trainable parameters, given the single critic shared across all agents in the system. Yet this algorithmic advantage comes with several drawbacks. First, it requires information and parameter sharing across agents, a particularly notable weakness in competitive environments where no information should be shared between competitors, as is the case in electricity markets. Second, from an implementation standpoint, the memory requirements of centralized critics grow substantially with the number of agents, constraining the implementation and the selection of conservative hyperparameters, which have proven crucial for multi-agent environments \cite{yu_surprising_2022}. Finally, from an empirical perspective, the centralized critic in MAPPO has shown no substantial performance advantage over independent learning algorithms in several benchmarks \cite{de_witt_is_2020}.

Considering these advantages and disadvantages, together with the work of \citet{gonzalez-ruiz_assessing_2026}, we select IPPO for the main simulations of this work, using conservative hyperparameters to partially mitigate the non-stationarity, while comparing the outcomes of both algorithms in the supplementary material. League-based training configurations, such as those implemented in \citet{vinyals_grandmaster_2019}, mitigate issues with independent learning, but we consider their computational cost beyond the scope of our current academic analysis.

Both algorithms use the RLlib framework \cite{liang_rllib_2018}. For IPPO, we use the implementation provided directly by the library, whereas for MAPPO we follow a public submission to the repository at \citet{httpsgithubcommatthewcweston_rllib_2025} and adapt it to our framework. The public repository, to be shared after acceptance of this manuscript, will include both implementations and runnable scripts.

\clearpage

\section{Scenarios}

This section describes complementary information for the scenarios presented in the main work, starting with the configuration of the GENCOs and the policymaker. The two agent types discount future cash flows at different rates, reflecting their distinct opportunity costs of capital: GENCOs apply an annual rate of $r_{g} = 8\%$, consistent with private investors' return requirements, while the policymaker uses a lower social discount rate of $r_{p} = 5\%$. The sixteen GENCOs are further divided into incumbents and entrants. Incumbent GENCOs are allocated the installed capacity reported in Table~\ref{Appendix table: Installed capacity 2025}, yielding a Herfindahl Hirschman Index (HHI) of 1{,}388. We acknowledge that this value exceeds current concentration levels in the Italian generation market, where leading producers hold shares well below those implied by this index \citep{arera_annual_2025}, but it provides a deliberate representation of a concentrated market.

\begin{table}[hbt!]
\centering
\begin{tabular}{@{}llll@{}}
\toprule
\multicolumn{4}{c}{\textbf{\begin{tabular}[c]{@{}c@{}}Installed capacity 2025\\ {[}MW{]}\end{tabular}}} \\ \midrule
Solar            & 40,294 & ESS 3-hours & 6,500  \\
Onshore wind     & 13,325 & ESS 8-hours & 0      \\
Offshore Wind    & 0      & Coal        & 5,326  \\
Hydro reservoir  & 11,766 & OCGT        & 8,500  \\
Run-of-the river & 9,571  & CCGT        & 45,000 \\
Others           & 8,500  &             &       
\end{tabular}%
\caption{\textbf{Installed capacity in 2025 for all scenarios}. Based on \citet{terna_terna_2026}.}
\label{Appendix table: Installed capacity 2025}
\end{table}

Regarding generation technologies, the model includes a selection of key assets for the energy transition, with detailed characteristics reported in Table~\ref{Appendix table: Generation Technology characteristics}. Although this set is smaller than those in established planning models, it captures the sector's main trends. In addition, it could be extended to include long-duration storage options as part of future work.

\begin{table}[hbt!]
\centering
\resizebox{\textwidth}{!}{%
\begin{tabular}{ccccccccc}
\hline
\textbf{\begin{tabular}[c]{@{}c@{}}Feature  \\ Technology\end{tabular}} &
  \textbf{\begin{tabular}[c]{@{}c@{}}Solar\\ PV\end{tabular}} &
  \textbf{\begin{tabular}[c]{@{}c@{}}Onshore \\ Wind\end{tabular}} &
  \textbf{\begin{tabular}[c]{@{}c@{}}Offshore\\ Wind\end{tabular}} &
  \textbf{Coal} &
  \textbf{\begin{tabular}[c]{@{}c@{}}Open Cycle \\ Gas Turbine\end{tabular}} &
  \textbf{\begin{tabular}[c]{@{}c@{}}Combined Cycle\\ Gas Turbine\end{tabular}} &
  \textbf{\begin{tabular}[c]{@{}c@{}}ESS\\ 3-hours\end{tabular}} &
  \textbf{\begin{tabular}[c]{@{}c@{}}ESS\\ 8-hours\end{tabular}} \\ \hline
\begin{tabular}[c]{@{}c@{}}CAPEX\\ 2025-2040  \\ {[}MW{]}\end{tabular} &
  594-385 &
  1438-1296 &
  2371-2179 &
  4812-4812 &
  593-557 &
  1142-1078 &
  1037-445 &
  2287-1010 \\ \hline
\begin{tabular}[c]{@{}c@{}}OPEX - Fixed \\ {[}\% CAPEX\\ MW - year{]}\end{tabular} &
  2.2 &
  1.7 &
  2.3 &
  1.31 &
  1.77 &
  3.3 &
  0.25 &
  0.25 \\ \hline
\begin{tabular}[c]{@{}c@{}}OPEX - Variable\\ {[}\$/MWh{]}\end{tabular} &
  1 &
  1.5 &
  1.5 &
  N/A &
  N/A &
  N/A &
  N/A &
  N/A \\ \hline
\begin{tabular}[c]{@{}c@{}}Emission factor \\ {[}CO2/MWh{]}\end{tabular} &
  0 &
  0 &
  0 &
  0.944 &
  0.347 &
  0.489 &
  N/A &
  N/A \\ \hline
\begin{tabular}[c]{@{}c@{}}Construction\\ time \\ {[}\# - years{]}\end{tabular} &
  2 &
  3 &
  4 &
  3 &
  3 &
  2 &
  2 &
  2 \\ \hline
\begin{tabular}[c]{@{}c@{}}Investment\\ steps\\ {[}MW{]}\end{tabular} &
  \multicolumn{8}{c}{[0,50,125,375,1125,3375{]}} \\ \hline
\begin{tabular}[c]{@{}c@{}}Operational\\ Lifetime\\ {[}\# - years{]}\end{tabular} &
  35 &
  28 &
  30 &
  40 &
  25 &
  25 &
  20 &
  20 \\ \hline
\begin{tabular}[c]{@{}c@{}}Expected\\ availability \\ {[}\%{]}\end{tabular} &
  90 &
  90 &
  90 &
  90 &
  90 &
  90 &
  90 &
  90 \\ \hline
\end{tabular}%
}
\caption{\textbf{Characteristics and relevant parameters for Generation and Storage technologies}. Based on \citet{international_energy_agency_average_2019,gumber_global_2024}}
\label{Appendix table: Generation Technology characteristics}
\end{table}

Natural gas prices are subject to occasional but persistent crisis episodes that a deterministic baseline trajectory does not capture. To represent these, we superimpose a stochastic shock process on the baseline gas price through a three-state Markov regime-switching model operating at the bimonthly resolution of the environment. At each step the chain occupies one of three states (\textit{Normal}, \textit{Shock}, or \textit{Recovery}) and returns a multiplier $m_t$ that scales the baseline gas-indexed variable cost. In the \textit{Normal} state, the multiplier is unity, and a shock arrives with per-bimester probability $\lambda = 0.017$. Once the time spent in elevated states is accounted for, this corresponds to roughly one crisis every 12 years. When a shock arrives, the chain jumps to the \textit{Shock} state and the multiplier is set to a peak level $m^{peak}$ drawn from a log-normal distribution, $m^{peak} \sim \mathrm{LogNormal}(\ln 2,\, 0.25)$ clipped to $[1.3, 3.5]$, so that the median peak doubles the long-term price while retaining a realistic dispersion.

The system then remains in the \textit{Shock} state with per-bimester persistence $p_{stay} = 11/12$, implying an expected crisis duration of $1/(1 - p_{stay}) = 12$ bimesters (about two years). On exiting, the chain passes through a single \textit{Recovery} bimester, in which the multiplier is set to the midpoint between the peak and unity, before returning to \textit{Normal}. The module can also be deactivated over horizons where gas prices are pinned to an exogenous source \textit{(e.g. the observed forward curve during the simulation warm-up)}: while inactive it returns a unit multiplier and holds the chain in \textit{Normal}, preventing latent transitions from surfacing when activation is later restored. 

The baseline gas price for 2025-2030 is taken from the PyPSA forward curve, which already reflects the currently elevated price levels. Thus, the stochastic shock module is therefore activated only from 2030 onward, once the forward curve reverts to long-term levels, so that the present crisis is not double-counted.

\begin{figure}[hbt!]
\centering
\includegraphics[width=1.0\linewidth]{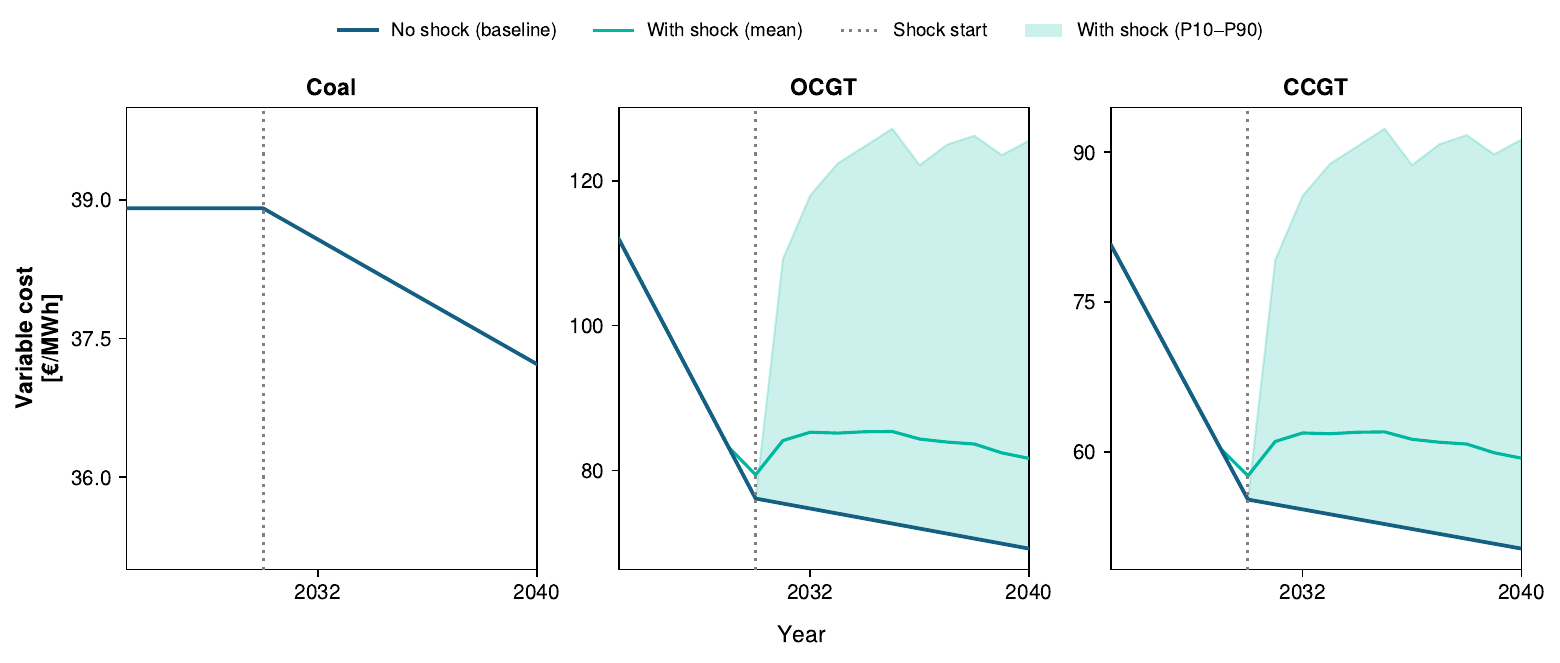}
\caption{\textbf{Stochastic natural gas price shocks and their effect on thermal generation costs.} One representative realization of the gas-price shock process over the simulation horizon, showing the resulting variable generation costs of coal, CCGT, and OCGT plants.}
\label{Appendix fig: Natural gas shocks}
\end{figure}

\clearpage

\section{Experiments}

Across experiments, we employ the hyperparameter configuration detailed in Table~\ref{Appendix table: Hyperparameters}. The selection follows findings from previous multi-agent PPO implementations \citep{yu_surprising_2022,gonzalez-ruiz_assessing_2026,de_witt_is_2020}, which adopt a conservative approach that trades learning throughput for training stability, reflected here in large batch sizes, the small clipping parameter, and the low learning rate. We note that the discount factor is set to $\gamma = 1$, as discounting is handled within the environment using agent-specific rates, avoiding double discounting. We apply a uniform hyperparameter configuration to all agents in the system. Future research could explore more extensive hyperparameter optimization for this market environment, including the potential benefits of differentiated settings across distinct agent categories.

\begin{table}[!hbt]
\centering
\resizebox{\textwidth}{!}{%
\begin{tabular}{@{}llll@{}}
\toprule
\textbf{Hyperparameter}               & \textbf{Value} & \textbf{Hyperparameter}   & \textbf{Value} \\ \midrule
Batch Size                            & 10800          & Parallel Sampling Workers & 60             \\
Mini-batch size                       & 1800           & Environment per workers   & 1              \\
Stochastic gradient ascent iterations & 10             & Discount Factor - \( \gamma\)                     & 1              \\
Clipping parameter                    & 0.1            & Discount factor in Generalized Advantage Estimation - \( \lambda\)                    & 0.995          \\
Entropy Coefficient                   & 0.0            & kl coefficient            & 0.2            \\
Learning Rate                         & 5.00E-05       & kl target                 & 0.05          
\end{tabular}%
}
\caption{\textbf{List of hyperparameters.} Notation is kept consistent with the definitions in the RLlib Library.}
\label{Appendix table: Hyperparameters}
\end{table}

Table~\ref{Appendix table: Network architecture} reports the network configurations used across experiments. In IPPO, both the actor and the critic are simple MLPs, instantiated independently for each agent. In MAPPO, the actor is analogous to that of IPPO, but the critic uses a more elaborate architecture to handle the larger volume of information arising from the centralized observation set. An agent encoder first reduces the dimensionality of each agent's observations, a central mixer then combines the aggregated information, and finally, it feeds it to per-agent value-function heads. This design, together with the value normalization proposed by \citet{yu_surprising_2022}, allows the network to learn value representations per agent and prevents any single agent from dominating the learned value function.

\begin{table}[!hbt]
\centering
\begin{tabular}{@{}llll@{}}
\toprule
\textbf{IPPO Actor/Critic} & \textbf{Value} & \textbf{MAPPO Critic} & \textbf{Value} \\ \midrule
Enconder                   & {[}512,512{]}  & Agent Encoder         & {[}256,32,8{]} \\
Action head                & {[}64{]}       & Central Mixer         & {[}256,256{]}  \\
Value-function head        & {[}64{]}       & Value-function head   & {[}64{]}       \\
Activation layers          & RELU           & Activation layers     & RELU           \\ \bottomrule
\end{tabular}%
\caption{\textbf{Network architectures for IPPO and MAPPO experiments.}}
\label{Appendix table: Network architecture}
\end{table}

Last, the implementation uses Python 3.9.18, Ray 2.51.2, PyTorch 2.6.0, and CUDA 12.6. For training, we use a single computing node from a supercomputing system, each consisting of dual Intel Xeon Max 9480 processors with 56 cores \textit{(61 allocated for training)}, 1 TB of RAM \textit{(150 GB allocated for training)}, and 8 Nvidia H100 GPUs \textit{(1 GPU from the H100 cluster used for training)}. We cap each training session at 24 hours, a budget consistent with both the shared use of the supercomputing system and the convergence to an equilibrium consistently observed during training. Sampling is performed using a single computing core and 8 GB of RAM, yielding a single environment trajectory in approximately one minute. 

\clearpage

\section{Independent versus Multi-Agent Proximal Policy Optimization}

This section serves a double purpose. First, it provides a general robustness analysis of the methods, showing that the main conclusions of this work are reliably recovered from the solution approach. Second, it compares IPPO and MAPPO as MARL algorithms in the current setup. To this end, we train the \textbf{[H]} scenario for both the \textbf{[ND]} and \textbf{[D]} policy cases, using three seeds each. The training framework is replicated under comparable setups for IPPO and MAPPO, maintaining the same computational budget as in the main experiments.

\begin{figure}[hbt!]
\centering
\includegraphics[width=1.0\linewidth]{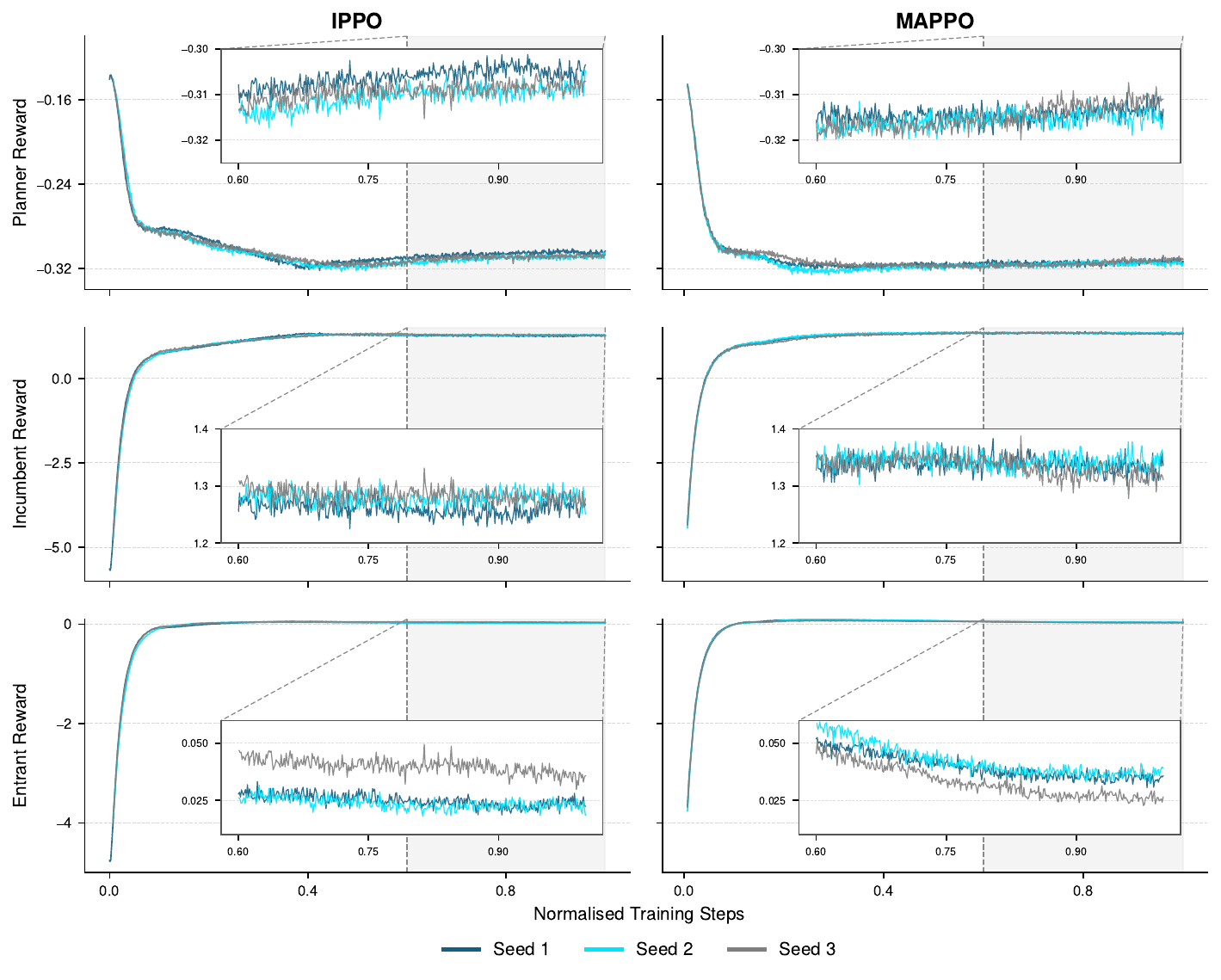}
\caption{\textbf{Evolution of mean agent reward by category during training in the \textbf{[H]} and \textbf{[ND]} scenario using IPPO and MAPPO solution algorithms.} Training steps are normalized by the maximum value reached within the computational budget. Agents are grouped into the policymaker, incumbent GENCOs, and entrant GENCOs. The inset highlights agent behavior in the final stages of training. All axis limits are shared to facilitate visualization.}
\label{Appendix fig: Training IPPO MAPPO ND}
\end{figure}

\begin{figure}[hbt!]
\centering
\includegraphics[width=1.0\linewidth]{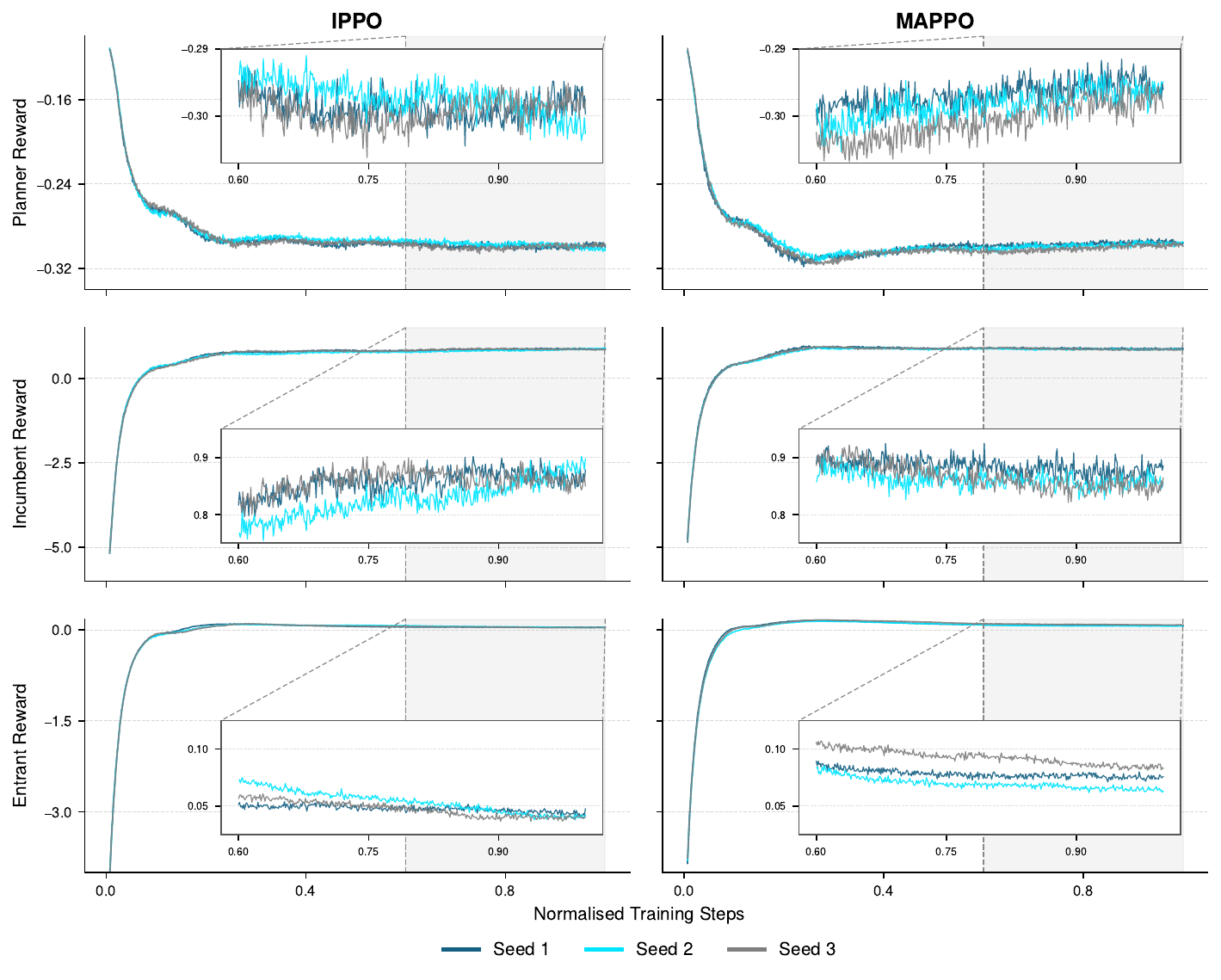}
\caption{\textbf{Evolution of mean agent reward by category during training in the \textbf{[H]} and \textbf{[D]} scenario using IPPO and MAPPO solution algorithms.} Training steps are normalized by the maximum value reached within the computational budget. Agents are grouped into the policymaker, incumbent GENCOs, and entrant GENCOs. The inset highlights agent behavior in the final stages of training. All axis limits are shared to facilitate visualization.}
\label{Appendix fig: Training IPPO MAPPO D}
\end{figure}

Figures \ref{Appendix fig: Training IPPO MAPPO ND} and \ref{Appendix fig: Training IPPO MAPPO D} present the training curves across these experiments. Three results are worth highlighting. First, across seeds, the training profiles are highly aligned, with relatively small differences in rewards between agents. Second, the training profiles of IPPO and MAPPO follow similar trends. Finally, these previous trends hold for the \textbf{[ND]} and \textbf{[D]} scenarios. In short, although relative variations are observed, the training profiles remain consistent across methods and scenarios.

\begin{figure}[hbt!]
\centering
\includegraphics[width=1.0\linewidth]{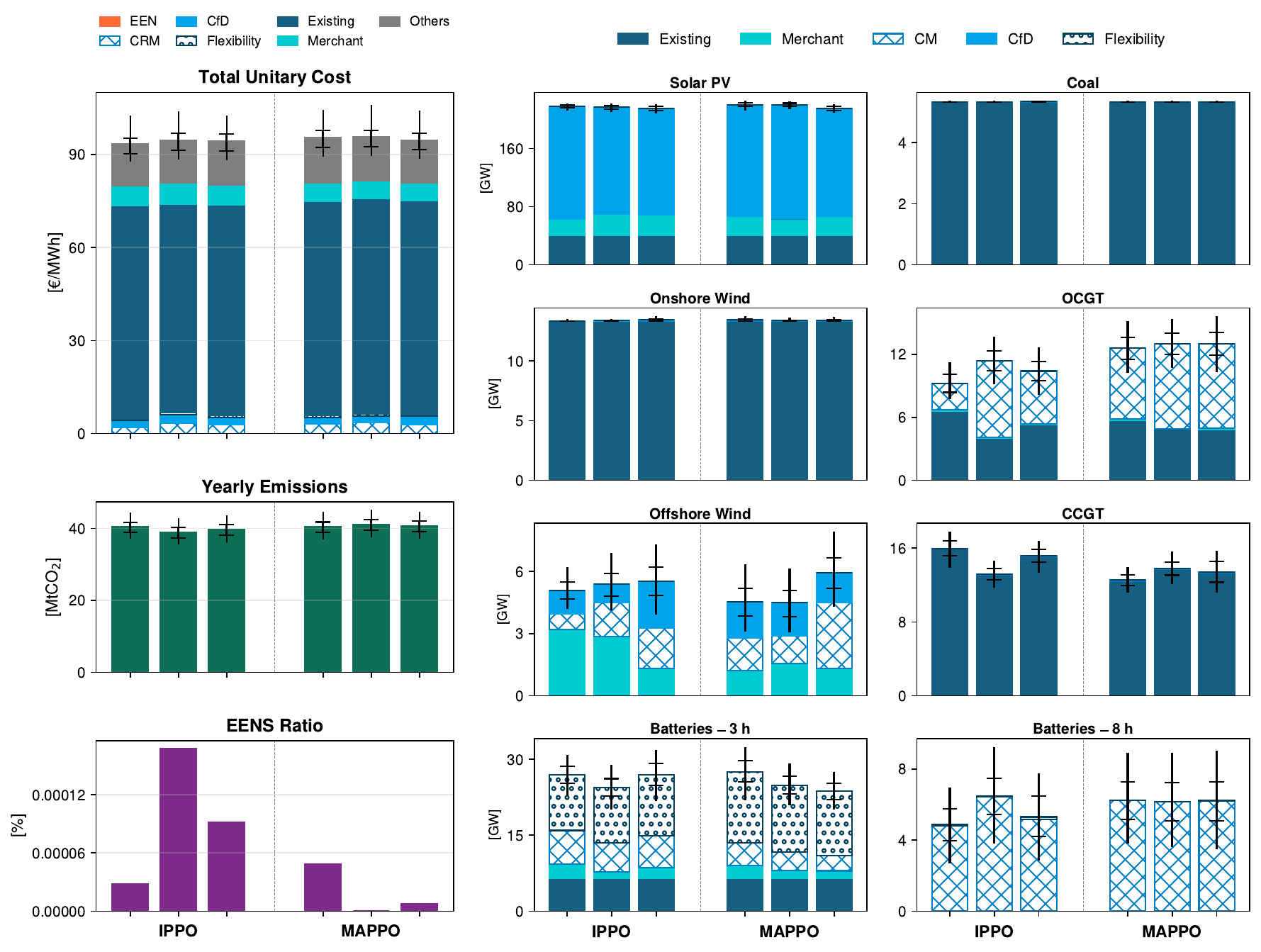}
\caption{\textbf{Total Unitary costs, CO$_2$ emissions, Energy not supplied, and Installed capacities in 2040 in the \textbf{[H]} and \textbf{[ND]} scenario using IPPO and MAPPO solution algorithms.} In each bar, the horizontal markers display the 25th and 75th percentiles, and the vertical markers display the 5th and 95th percentiles. For both total unitary cost bars and installed capacities, hatching patterns indicate the contribution of each market mechanism to the final cost: the Contracts for Differences, Capacity Market, and Flexibility components reflect the financial settlement of their respective mechanisms; Other captures the wholesale remuneration of capacity and flexibility assets, which remain exposed to the short-term price signal; and Existing and Merchant bars refer to the wholesale market remuneration of old and new assets.}
\label{Appendix fig: Composite ND}
\end{figure}

\begin{figure}[hbt!]
\centering
\includegraphics[width=1.0\linewidth]{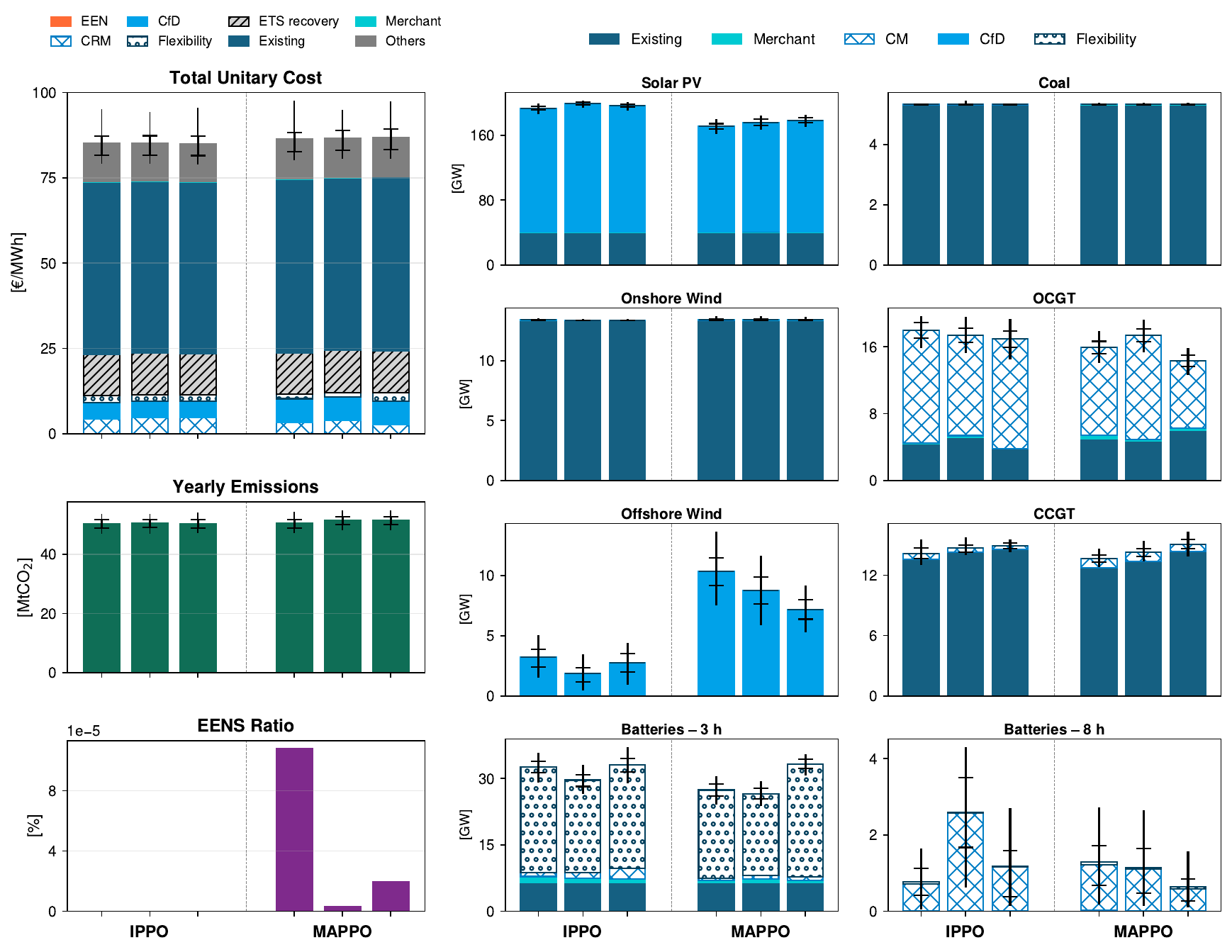}
\caption{\textbf{Total Unitary costs, CO$_2$ emissions, Energy not supplied, and Installed capacities in 2040 in the \textbf{[H]} and \textbf{[D]} scenario using IPPO and MAPPO solution algorithms.} In each bar, the horizontal markers display the 25th and 75th percentiles, and the vertical markers display the 5th and 95th percentiles. For both total unitary cost bars and installed capacities, hatching patterns indicate the contribution of each market mechanism to the final cost: the Contracts for Differences, Capacity Market, and Flexibility components reflect the financial settlement of their respective mechanisms; Other captures the wholesale remuneration of capacity and flexibility assets, which remain exposed to the short-term price signal; Existing and Merchant bars refer to the wholesale market remuneration of old and new assets; and ETS recovery represents the ETS costs recovered through the tariff.}
\label{Appendix fig: Composite D}
\end{figure}

Once the training budget is exhausted, aggregate market results are obtained and shown in Figures \ref{Appendix fig: Composite ND} and \ref{Appendix fig: Composite D}. As with the training profiles, market outcomes are consistent across seeds and methods. Moreover, the central claims of the main work are preserved: applying the \textit{Decreto Bollette} yields a slight cost reduction, substantially increases emissions, and substantially reduces merchant investment. Some changes in the generation mix do emerge across methods. For instance, IPPO seeds lead to greater uptake of solar PV entering the market through CfD auctions, which is replaced by offshore wind in the MAPPO case. Technologies are also interchanged across seeds: among ESS, some seeds favor 3-hour batteries while others favor 8-hour ones. These variations are not large enough to affect aggregate system cost and emissions, which remain largely aligned across the scenario matrix.

This analysis supports the suitability of IPPO, which attains results comparable to MAPPO without requiring a centralized critic. We note that MAPPO presents scalability issues in our setup, as the RAM and GPU memory requirements of centralized-critic training do not scale well with the number of agents, a limitation that IPPO avoids. While this is not a major constraint for the present work and could be addressed through environmental or algorithmic modifications, it reinforces IPPO as a simpler and more effective choice for our setup and application. Most importantly, this section underscores that the results of the main paper are robust across methodologies and seeds, besides the small variations emerging during training. 

\clearpage

\section{Additional information and results}

This section introduces additional information and results, complementing the main document. To start, Table \ref{Appendix table: mechanism capacity} summarizes the awarded capacity under the three main long-term mechanisms currently implemented in the Italian electricity system, highlighting the relevant technologies for the current analysis. 

\begin{table}[!hbt]
\centering
\resizebox{\textwidth}{!}{%
\begin{tabular}{@{}lllllll@{}}
\toprule
\multirow{2}{*}{\textbf{Year}} &
  \multicolumn{2}{c}{\textbf{FER - FERX}} &
  \multicolumn{3}{c}{\textbf{CRM}} &
  \multicolumn{1}{c}{\textbf{MACSE}} \\
 &
  \multicolumn{1}{c}{Solar PV} &
  \multicolumn{1}{c}{Onshore wind} &
  \multicolumn{1}{c}{Gas} &
  \multicolumn{1}{c}{Storage} &
  \multicolumn{1}{c}{Others} &
  \multicolumn{1}{c}{Batteries} \\ \midrule
\textbf{2022} & 1.30  & 0.12 & 2.58 & 1.13 & 0.09 & 0    \\
\textbf{2023} & 1.95  & 0.66 & 2.67 & 1.22 & 0.09 & 0    \\
\textbf{2024} & 2.41  & 0.75 & 2.71 & 1.30 & 0.10 & 0    \\
\textbf{2025} & 10.39 & 1.78 & 2.74 & 1.86 & 0.10 & 1.25 \\ \bottomrule
\end{tabular}%
}
\caption{\textbf{Cumulative awarded capacity [GW] in Italy's main renewable-support 
and capacity-adequacy mechanisms for the 2022-2025 horizon.} Renewable volumes are nominal 
installed capacity awarded under the FER schemes; figures for 2022-2024 correspond 
to legacy FER auctions, while the value for 2025 aggregates both the last iteration of the FER scheme and the first FER~X transitional auction (results published December~2025). Capacity Remuneration Mechanism (CRM) values report new-entrant capacity only, excluding existing capacity. MACSE reports the power rating of awards from the first auction (held September~2025). Compiled by the authors from 
\cite{terna_capacity_2026,sergio_italys_2025,terna_terna_2025}.}
\label{Appendix table: mechanism capacity}
\end{table}

Next, Tables \ref{Appendix table: Difference statistical analysis - Costs} and \ref{Appendix table: Difference statistical analysis - Emissions} complement the difference of costs and emissions presented in the main text, by comparing the obtained distributions from the simulation. The procedure carried out is the following. For each market configuration and time window, we test whether suppressing the carbon price signal (\textbf{[D]}) shifts total system costs and CO$_2$ emissions relative to the counterfactual (\textbf{[ND]}). Because each scenario is an independent training session with no shared random seeds, the two samples are treated as unpaired. The unit of analysis is the individual simulation: each of the $\sim$1{,}000 Monte Carlo runs per scenario is reduced to a single scalar, the demand-weighted mean unit cost and the annualized total emissions over the window, yielding an empirical outcome distribution per scenario. 

\begin{table}[!hbt]
\centering
\resizebox{\textwidth}{!}{%
\begin{tabular}{@{}clcccc@{}}
\toprule
\textbf{Scenario} &
  \multicolumn{1}{c}{\textbf{\begin{tabular}[c]{@{}c@{}}Time \\ Window\end{tabular}}} &
  \textbf{\begin{tabular}[c]{@{}c@{}}Percentage Difference\\ and 95\% confidence \\ interval {[}\%{]}\end{tabular}} &
  \textbf{\begin{tabular}[c]{@{}c@{}}Effect Size\\ (Hedges' g)\end{tabular}} &
  \textbf{Magnitude} &
  \textbf{Assessment} \\ \midrule
\multirow{4}{*}{\textbf{H+}}  & Aggregate & \begin{tabular}[c]{@{}c@{}}-10.6 \\ {[}-10.86, -10.34{]}\end{tabular} & -3.39  & Large  & Significant \\
                              & Short     & \begin{tabular}[c]{@{}c@{}}-15.1\\ {[}-15.15, -15.05{]}\end{tabular}  & -26.47 & Large  & Significant \\
                              & Mid       & \begin{tabular}[c]{@{}c@{}}-14.96\\ {[}-15.35, -14.55{]}\end{tabular} & -2.99  & Large  & Significant \\
                              & Long      & \begin{tabular}[c]{@{}c@{}}-1.62\\ {[}-1.99, -1.24{]}\end{tabular}    & -0.37  & Small  & Significant \\ \midrule
\multirow{4}{*}{\textbf{H}}   & Aggregate & \begin{tabular}[c]{@{}c@{}}-8.8\\ {[}-9.22, -8.39{]}\end{tabular}     & -1.8   & Large  & Significant \\
                              & Short     & \begin{tabular}[c]{@{}c@{}}-14.94\\ {[}-14.99, -14.89{]}\end{tabular} & -25.62 & Large  & Significant \\
                              & Mid       & \begin{tabular}[c]{@{}c@{}}-11.79\\ {[}-12.27, -11.31{]}\end{tabular} & -2.04  & Large  & Significant \\
                              & Long      & \begin{tabular}[c]{@{}c@{}}-1.96\\ {[}-2.8, -1.12{]}\end{tabular}     & -0.2   & Small  & Significant \\ \midrule
\multirow{4}{*}{\textbf{H-}}  & Aggregate & \begin{tabular}[c]{@{}c@{}}-10.74\\ {[}-11.22, -10.25{]}\end{tabular} & -1.86  & Large  & Significant \\
                              & Short     & \begin{tabular}[c]{@{}c@{}}-15.01\\ {[}-15.06, -14.96{]}\end{tabular} & -27.07 & Large  & Significant \\
                              & Mid       & \begin{tabular}[c]{@{}c@{}}-11.27\\ {[}-11.82, -10.68{]}\end{tabular} & -1.63  & Large  & Significant \\
                              & Long      & \begin{tabular}[c]{@{}c@{}}-8.31\\ {[}-9.25, -7.36{]}\end{tabular}    & -0.74  & Medium & Significant \\ \midrule
\multirow{4}{*}{\textbf{CRM}} & Aggregate & \begin{tabular}[c]{@{}c@{}}-11.94\\ {[}-12.54, -11.31{]}\end{tabular} & -1.64  & Large  & Significant \\
                              & Short     & \begin{tabular}[c]{@{}c@{}}-14.98\\ {[}-15.03, -14.93{]}\end{tabular} & -25.57 & Large  & Significant \\
                              & Mid       & \begin{tabular}[c]{@{}c@{}}-14.37\\ {[}-15.01, -13.71{]}\end{tabular} & -1.8   & Large  & Significant \\
                              & Long      & \begin{tabular}[c]{@{}c@{}}-7.58\\ {[}-8.76, -6.38{]}\end{tabular}    & -0.54  & Medium & Significant \\ \bottomrule
\end{tabular}%
}
\caption{\textbf{Statistical significance of the difference between [ND] and [D] scenarios on the total system across time horizons.} Each row reports the percentage difference between the \textbf{[D]} and \textbf{[ND]} scenarios, $(\mathrm{D}-\mathrm{ND})/\mathrm{ND}$, with its 95\% bootstrap confidence interval, the effect size (Hedges' $g$) with its magnitude band. Given the large number of simulations per scenario, the effect size and the confidence interval carry the interpretation; the assessment column flags differences that are statistically significant yet of negligible magnitude.}
\label{Appendix table: Difference statistical analysis - Costs}
\end{table}

\begin{table}[!hbt]
\centering
\resizebox{\textwidth}{!}{%
\begin{tabular}{@{}clcccc@{}}
\toprule
\textbf{Scenario} &
  \multicolumn{1}{c}{\textbf{\begin{tabular}[c]{@{}c@{}}Time \\ Window\end{tabular}}} &
  \textbf{\begin{tabular}[c]{@{}c@{}}Percentage Difference\\ and 95\% confidence \\ interval {[}\%{]}\end{tabular}} &
  \textbf{\begin{tabular}[c]{@{}c@{}}Effect Size\\ (Hedges' g)\end{tabular}} &
  \textbf{Magnitude} &
  \textbf{Assessment} \\ \midrule
\multirow{4}{*}{\textbf{H+}}  & Aggregate & \begin{tabular}[c]{@{}c@{}}8.16 \\ {[}7.66, 8.65{]}\end{tabular}    & 1.5   & Large  & Significant \\
                              & Short     & \begin{tabular}[c]{@{}c@{}}3.23\\ {[}2.63, 3.85{]}\end{tabular}     & 0.47  & Small  & Significant \\
                              & Mid       & \begin{tabular}[c]{@{}c@{}}8.68\\ {[}8.07, 9.29{]}\end{tabular}     & 1.32  & Large  & Significant \\
                              & Long      & \begin{tabular}[c]{@{}c@{}}15.57\\ {[}13.21, 18.01{]}\end{tabular}  & 0.62  & Medium & Significant \\ \midrule
\multirow{4}{*}{\textbf{H}}   & Aggregate & \begin{tabular}[c]{@{}c@{}}24.6\\ {[}24.08, 25.12{]}\end{tabular}   & 4.66  & Large  & Significant \\
                              & Short     & \begin{tabular}[c]{@{}c@{}}6.76\\ {[}6,12, 7.41{]}\end{tabular}     & 0.96  & Large  & Significant \\
                              & Mid       & \begin{tabular}[c]{@{}c@{}}26.51\\ {[}25.83, 27.18{]}\end{tabular}  & 3.95  & Large  & Significant \\
                              & Long      & \begin{tabular}[c]{@{}c@{}}33.57\\ {[}32.28, 34.91{]}\end{tabular}  & 2.61  & Large  & Significant \\ \midrule
\multirow{4}{*}{\textbf{H-}}  & Aggregate & \begin{tabular}[c]{@{}c@{}}35.05\\ {[}34.48, 35.61{]}\end{tabular}  & 6.62  & Large  & Significant \\
                              & Short     & \begin{tabular}[c]{@{}c@{}}5.92\\ {[}5.29, 6.57{]}\end{tabular}     & 0.83  & Large  & Significant \\
                              & Mid       & \begin{tabular}[c]{@{}c@{}}39.65\\ {[}38.88, 40.45{]}\end{tabular}  & 5.5   & Large  & Significant \\
                              & Long      & \begin{tabular}[c]{@{}c@{}}45.25\\ {[}44.01, 46.51{]}\end{tabular}  & 4.06  & Large  & Significant \\ \midrule
\multirow{4}{*}{\textbf{CRM}} & Aggregate & \begin{tabular}[c]{@{}c@{}}56.92\\ {[}56.28, 57.56{]}\end{tabular}  & 10.67 & Large  & Significant \\
                              & Short     & \begin{tabular}[c]{@{}c@{}}7.1\\ {[}6.46, 7.77{]}\end{tabular}      & 0.97  & Large  & Significant \\
                              & Mid       & \begin{tabular}[c]{@{}c@{}}49.88\\ {[}49.12, 50.65{]}\end{tabular}  & 7.45  & Large  & Significant \\
                              & Long      & \begin{tabular}[c]{@{}c@{}}98.83\\ {[}97.13, 100.52{]}\end{tabular} & 8.39  & Large  & Significant \\ \bottomrule
\end{tabular}%
}
\caption{\textbf{Statistical significance of the difference between [ND] and [D] scenarios on CO$_2$ emissions across scenarios and time horizon.} Each row reports the percentage difference between the \textbf{[D]} and \textbf{[ND]} scenarios, $(\mathrm{D}-\mathrm{ND})/\mathrm{ND}$, with its 95\% bootstrap confidence interval, the effect size (Hedges' $g$) with its magnitude band. Given the large number of simulations per scenario, the effect size and the confidence interval carry the interpretation; the assessment column flags differences that are statistically significant yet of negligible magnitude.}
\label{Appendix table: Difference statistical analysis - Emissions}
\end{table}

To assess differences in distributional location, we use the Welch's $t$-test and the Mann-Whitney $U$ test; we report the latter, which makes no distributional assumptions, and confirm that the two agree throughout. The difference in means is reported in physical units with a 95\% percentile bootstrap confidence interval ($10^4$ resamples), and as a percentage of the \textbf{[ND]} baseline. To guard against false positives from testing four markets simultaneously, $p$-values are corrected within each metric and window using the Holm step-down procedure.

We lead the interpretation with effect size: Hedges' $g$ (the standardized mean difference, bias-corrected for sample size) and Cliff's $\delta$ (a rank-based measure of stochastic dominance), classified into negligible/small/medium/large bands following standard thresholds \cite{cohen_statistical_2009,meissel_using_2024}. Each comparison is summarized by a verdict that combines the corrected $p$-value with the effect-size band, distinguishing differences that are both significant and non-negligible from those that are statistically detectable yet practically negligible. 

Lastly, Table \ref{Appendix table: heatmap nominal values} presents comparative metrics across scenarios for nominal values, complementing the corresponding graph in the main text, which normalizes them for visualization purposes. 

\begin{table}[!hbt]
\centering
\resizebox{\textwidth}{!}{%
\begin{tabular}{lllllllllll}
\hline
\textbf{Market} &
  \textbf{Scenario} &
  \textbf{\begin{tabular}[c]{@{}l@{}}Wholesale \\ volatility\\ {[}-{]}\end{tabular}} &
  \textbf{\begin{tabular}[c]{@{}l@{}}Total Cost \\ volatility\\ {[}-{]}\end{tabular}} &
  \textbf{\begin{tabular}[c]{@{}l@{}}Gas shock \\ vulnerability\\ {[}\%{]}\end{tabular}} &
  \textbf{\begin{tabular}[c]{@{}l@{}}Profit \\ incumbents\\ {[}\${]}\end{tabular}} &
  \textbf{\begin{tabular}[c]{@{}l@{}}Profit\\ entrants\\ {[}\${]}\end{tabular}} &
  \textbf{\begin{tabular}[c]{@{}l@{}}Merchant \\ share\\ {[}\%{]}\end{tabular}} &
  \textbf{\begin{tabular}[c]{@{}l@{}}Existing \\ share\\ {[}\%{]}\end{tabular}} &
  \textbf{\begin{tabular}[c]{@{}l@{}}Mechanism\\ share\\ {[}\%{]}\end{tabular}} &
  \textbf{\begin{tabular}[c]{@{}l@{}}Energy not \\ served ratio\\ {[}\%{]}\end{tabular}} \\ \hline
\multirow{3}{*}{\textbf{H+}}  & ND & 1.1  & 0.19 & 2.8   & 1.4  & 0.12  & 2.4   & 60 & 23  & 0       \\
                              & UD & 0.94 & 0.2  & -0.72 & 1.1  & 0.17  & 0.082 & 55 & 24  & 0       \\
                              & D  & 0.94 & 0.16 & 3.2   & 0.81 & 0.15  & 0.03  & 46 & 34  & 0       \\ \hline
\multirow{3}{*}{\textbf{H}}   & ND & 0.47 & 0.29 & 1.6   & 1.7  & 0.1   & 6.7   & 74 & 5.3 & 0       \\
                              & UD & 0.37 & 0.23 & 5.5   & 1.7  & 0.17  & 3     & 71 & 4.5 & 0       \\
                              & D  & 0.37 & 0.18 & 3.4   & 1.2  & 0.13  & 0.056 & 59 & 14  & 0       \\ \hline
\multirow{3}{*}{\textbf{H-}}  & ND & 0.44 & 0.34 & 3.7   & 1.8  & 0.093 & 14    & 74 & 2.8 & 0       \\
                              & UD & 0.35 & 0.24 & 8.3   & 1.8  & 0.2   & 9.5   & 71 & 2.7 & 0       \\
                              & D  & 0.3  & 0.17 & 12    & 1.2  & 0.084 & 0.96  & 66 & 7.3 & 0       \\ \hline
\multirow{3}{*}{\textbf{CRM}} & ND & 0.5  & 0.46 & 5.3   & 1.9  & 0.14  & 21    & 75 & 2.1 & 0.001   \\
                              & UD & 0.27 & 0.25 & 5     & 1.8  & 0.2   & 19    & 74 & 3   & 0       \\
                              & D  & 0.35 & 0.26 & 7.8   & 1.3  & 0.059 & 7.7   & 70 & 1.5 & 0.00026 \\ \hline
\end{tabular}%
}
\caption{\textbf{Aggregate key metrics across the scenario matrix.} Wholesale and total price volatility are computed as the standard deviation of hourly prices, using normalized yearly values to discount inter-year variability. Shock vulnerability measures the relative cost between simulations with and without the natural gas price shock. Profit metrics report the discounted net present value accruing to each agent category. Market contributions quantify the relative weight of merchant investments, existing assets, and long-term regulatory mechanisms in the final cost. Energy not served measures the unmet demand ratio across the simulations.}
\label{Appendix table: heatmap nominal values}
\end{table}


\clearpage


\bibliography{references}

\bigskip




\end{document}